\documentclass[12pt]{article}
\usepackage{amsfonts,amsthm,amsmath,amssymb,upgreek,bm}
\usepackage{dsfont}
\usepackage[paper=letterpaper,margin=.85in]{geometry}
\usepackage{graphicx}
\usepackage{units}
\usepackage{float}
\usepackage[export]{adjustbox}
\usepackage{bbold}
\usepackage[shortlabels]{enumitem}

\usepackage{color}
\usepackage[dvipsnames]{xcolor}
\definecolor{darkblue}{rgb}{0.1,0.1,.7}
\definecolor{purple}{rgb}{0.6,0,0.6}
\definecolor{orange}{rgb}{0.9,0.6,0}
\usepackage[colorlinks, linkcolor=darkblue, citecolor=darkblue, urlcolor=darkblue, linktocpage]{hyperref} 
\usepackage[square, comma, compress,numbers]{natbib}
\usepackage{physics}
\usepackage{tensor}
\usepackage{psfrag}
\usepackage{footmisc}
\usepackage{url}
\usepackage{mathtools}
\usepackage{tikz}
    \usepackage{amssymb,amsfonts,amsmath}
    \usepackage{tkz-euclide}
        \usetikzlibrary{arrows,calc,patterns}
\usepackage{pgfplots}
\usepackage[]{amsmath}
\usepackage[]{graphicx}
\usepackage[]{latexsym}
\usepackage[utf8]{inputenc}
\usepackage{geometry}
\usepackage{amscd}
\usepackage[all,cmtip]{xy}
\usepackage{mathrsfs}
\usepackage[most]{tcolorbox}
\usepackage{tikz}
\usepackage{subcaption}
\usetikzlibrary{arrows,positioning}

\def\SL2{\widetilde{SL}(2,\mathbb R)}

\def\hor{\text{hor}}

\numberwithin{equation}{section}
\newcommand{\be}{\begin{equation}}
\newcommand{\ee}{\end{equation}}

\renewcommand{\abs}[1]{\left\lvert #1 \right\rvert}

\newcommand {\bes} {\begin {equation*}}
\newcommand {\ees} {\end {equation*}}
\newcommand {\beq} {\begin {equation}}
\newcommand {\eeq} {\end {equation}}
\newcommand {\bea} {\begin {eqnarray}}
\newcommand {\ea} {\end {eqnarray}}
\newcommand {\eea} {\end {eqnarray}}

\numberwithin{equation}{section}

\renewcommand{\tilde}{\widetilde}
\newcommand{\xvec}{\bold{x}} 
\newcommand{\txvec}{\widetilde{\bold{x}}}

\renewcommand{\i}{{\rm i}}

\newcommand{\bast}{\boldsymbol{\ast}}

\newcommand{\tgamma}{\widetilde{\gamma}}

\newcommand{\barZ}{\overline{Z}} 
\newcommand{\barOmega}{\overline{\Omega}} 
\newcommand{\onebar}{{\bar{1}}}
\newcommand{\twobar}{{\bar{2}}}
\newcommand{\htilde}{\tilde{h}}
\renewcommand{\=}{\;= \;}

\newcommand{\Gidi}{{\langle \Gamma_{1} ,\delta_{1} \rangle}} 
\newcommand{\Gzdz}{\langle \Gamma_2 ,\delta_2 \rangle}

\newcommand{\didz}{\langle \delta_1 ,\delta_2 \rangle} 
\newcommand{\GiGz}{\langle \Gamma_1 ,\Gamma_2 \rangle} 
\newcommand{\hGi}{\langle h,\Gamma_1\rangle} 
 
\newcommand{\hdi}{\langle h ,\delta_1 \rangle} 
\newcommand{\hdz}{\langle h ,\delta_2 \rangle}

\newcommand{\CN}{\mathcal{N}}

\newcommand{\BH}{\text{BH}}

\newcommand{\omE}{\omega_E}

\newcommand{\Hv}{H}

\def\<{\langle}
\def\>{\rangle}

\tikzset{
    >=stealth',
    punkt/.style={
           rectangle,
           rounded corners,
           draw=black, very thick,
           text width=15em,
           minimum height=2em,
           text centered},
    pil/.style={
           ->,
           thick,
           shorten <=2pt,
           shorten >=2pt,}
}

\def\ie{\begin{equation}\begin{aligned}}
\def\fe{\end{aligned}\end{equation}}

\newcommand{\xgray}{x_{12}^{\ast}}

\begin{document}

\thispagestyle{empty}
\vspace*{2.5cm}
\begin{center}

{\LARGE\bfseries
    Multicentered black hole saddles for\\[1ex]
    supersymmetric indices
}
\begin{center}
\vspace{1cm}
 {\bf Jan Boruch${}^1$, Luca V. Iliesiu${}^1$,
Sameer Murthy${}^2$,
and Gustavo J. Turiaci${}^3$}\\
\bigskip \rm
\bigskip
${}^1$Center for Theoretical Physics and Department of Physics, University of California, Berkeley, California 94720, USA
\\
${}^2$Department of Mathematics, King's College London, The Strand, London WC2R 2LS, UK
\\
${}^3$Physics Department, University of Washington, Seattle, WA, USA
\rm
\end{center}
\vspace{2.5cm}
{\bf Abstract}
\end{center}

\begin{quotation}
\noindent
The supersymmetric index in string theory  can sometimes have a discontinuous integer-valued jump at co-dimension one surfaces in moduli space called walls of marginal stability. 
When the index counts black hole microstates, crossing such walls of marginal stability amounts to the appearance or disappearance of a large number of such states. 
While wall-crossing has been understood in string theory and through the disappearance of extremal Lorentzian supergravity solutions as the moduli are varied, there has been no understanding about how the discontinuous changes in the index occur at the level of the gravitational path integral. 
In this paper, we find the finite-temperature saddles in $4d$ flatspace supergravity in which fermionic fields are periodic when going around the thermal circle that correspond to the multi-center black hole contributions to the index. By analyzing these saddles, we can explain how wall-crossing occurs: as the scalar moduli in supergravity are varied at the asymptotic boundary, for a given split of the charges, the saddle point equations can no longer be solved and, consequently, the corresponding multi-center saddle no longer contributes to the index. While the values of the scalars and the jump in the index when a wall is crossed all agree with the prediction from previously found Lorentzian supergravity solutions, the saddles in the index exhibit a much richer moduli space, which we analyze in detail. 
\end{quotation}

\setcounter{page}{0}
\setcounter{tocdepth}{2}
\setcounter{footnote}{0}
\newpage

\tableofcontents
\newpage

\section{Introduction}

One of the great successes of string theory is in reproducing the Bekenstein-Hawking entropy as a microscopic count over states \cite{Strominger:1996sh}. This was done by computing a supersymmetric index, a quantity protected under changes of the couplings, in the effective theory living on the intersecting D-branes at small coupling. This index was then compared with the horizon area of the corresponding extremal supersymmetric black hole with matching charges. The results match to leading order, however, one could worry that we are essentially comparing two observables that are, at first sight, unrelated: an entropy which is supposed to capture the total degeneracy of the black hole microstates and an index which computes the degeneracy graded by the fermion number operator $(-1)^F$ of the contributing states. 
    
    To fully utilize the power of supersymmetric indices in order to learn about quantum gravity in regimes inaccessible 
    to weakly coupled string theory, 
one is motivated to define a notion of a gravitational supersymmetric index that can be computed directly using gravity techniques. 
Such an index can be defined by relying on the Gibbons-Hawking prescription \cite{Gibbons:1976ue}, which computes gravitational partition functions by summing over all possible geometries and field configurations consistent with asymptotic boundary conditions. 
To apply this procedure to compute supersymmetric indices, one considers an appropriate grand canonical partition function evaluated at complex values of chemical potentials, tuned exactly to values which turn the partition function into the supersymmetric index. For example, for black holes in flat space, one fixed the angular velocity $\Omega= 2\pi \i/\beta$, which is responsible for grading the degeneracy by the fermion number operator, $(-1)^F$.\footnote{Although the partition function is rotationally invariant for this choice of angular velocity,  the associated black hole geometry breaks this isometry by selecting an arbitrary axis of rotation with respect to which one defines the angular velocity. 
Moreover, the relevant partition function in flat space is not an index, but rather, a helicity supertrace. 
The difference between an index and a helicity supertrace is only important 
beyond the semi-classical level
of the gravitational path integral and will not concern us for the remainder of this paper.  }
    
    The crucial point of this approach is that, instead of working directly with supersymmetric black holes at zero temperature, one works with supersymmetric saddles of the index directly at finite temperature with periodic boundary conditions for all fermionic fields when going around the thermal circle. Such index saddles have been studied extensively in recent literature \cite{Cabo-Bizet:2018ehj,Cassani:2019mms,Bobev:2019zmz,Bobev:2020pjk,Larsen:2021wnu,Hristov:2021qsw,Hristov:2022pmo,BenettiGenolini:2023rkq,Iliesiu:2021are, H:2023qko,Anupam:2023yns,Boruch:2023gfn,Hegde:2023jmp,Chowdhury:2024ngg,Chen:2024gmc,Cassani:2024kjn,Hegde:2024bmb,Adhikari:2024zif,Heydeman:2024fgk,Boruch:2025qdq,Cassani:2025sim,BenettiGenolini:2025jwe,Bandyopadhyay:2025jbc} and shed a new light on many known supergravity solutions. The advantage of this approach is that it allows one to study gravitational indices without relying on the decoupling limit and the entropy function formalism \cite{Sen:2008vm,Dabholkar:2010rm,Dabholkar:2010uh,Iliesiu:2022kny,LopesCardoso:2022hvc}. 
    This opens up a new playground where one can study aspects of the gravitational path integral, whilst still being in setups where the comparison to exact microscopic answers  \cite{Strominger:1996sh,Maldacena:1999bp,Gaiotto:2005gf,Pioline:2005vi,Shih:2005qf,David:2006yn} is possible. 

    In this paper we construct a new family of supersymmetric finite temperature solutions of $\mathcal{N}=2$ supergravity coupled to vector multiplets in four-dimensional asymptotically flat space.
    These configurations are saddles of the gravitational index that capture the index of bound states of multiple black holes. 
    Hence we refer to these solutions as multicentered black hole saddles.

    Multicentered black holes at extremality \cite{Denef:2000nb,Bates:2003vx,Denef:2007vg} have already played an important role in the context of supersymmetric indices as they provide a physical picture for ``wall-crossing" -- a phenomenon in which a supersymmetric index can discretely jump as one varies the asymptotic boundary condition for the scalar fields. The codimension-one subregions of scalar moduli space at which the index jumps are referred to as "walls of marginal stability". When approaching the wall of marginal stability in moduli space, some of the black holes that form a bound state get further and further apart from each other, until finally some of them get separated by an infinite distance when one reaches the wall. Beyond that point, these black holes can no longer form a bound state, which is interpreted as a loss for the corresponding black hole microstates from Hilbert space; correspondingly, the supersymmetric index changes by a discrete value. While these multicentered black hole solutions match numerous known results for ``wall-crossing'' in the mathematical literature, they are not immune to the problems discussed above: they are only defined at zero temperature, and the sum of black hole areas is at most expected to capture a total entropy rather than a graded index.

    Rather, to construct the finite temperature supersymmetric index saddles, we follow the ``new attractor" approach presented in \cite{Boruch:2023gfn}. This allows us to write down a finite temperature supersymmetric index saddle for an arbitrary number of black holes. The solutions are specified in terms of positions of the north and south poles of the rotating black hole horizon (these poles are singled out by the choice of rotation axis), together with a choice of monopole and dipole charges of individual black holes in a given configuration. This data is constrained by smoothness and supersymmetry. This is sufficient to specify uniquely the solution in the presence of a single black hole \cite{Boruch:2023gfn}. In the presence of multiple horizons, the dipole charges are fixed in terms of the monopole charges, but a moduli space of solutions remains (arising from the relative locations of all the horizons and the choice of their axes of rotation). We show that the semiclassical contribution of these solutions to the index is always given simply by the sum of the corresponding extremal entropies of individual black holes -- the answer is exactly consistent with the general expectation that the gravitational index computes an index of some putative supersymmetric quantum mechanical system. The positions that describe a given configuration are subject to regularity conditions -- a set of algebraic equations which constrain the possible space of solutions in a way that ensures smoothness of the geometry. These conditions are a finite temperature generalization of the integrability conditions used in extremal Bates-Denef solutions \cite{Denef:2000nb}. We explain how the modification of these conditions changes the classical moduli space of multicentered solutions. In the simplest example of a two black hole bound state saddle, we show that the zero-temperature moduli space given by $S^2$ (rotations of the position of one center with respect to another) gets modified into an $S^1 \times S^3$ moduli space of solutions (combining both rotations of the black hole locations as well as the axis of rotation of each center). 
    As one passes through the wall of marginal stability, we find that the entire moduli space of finite temperature saddles vanishes, which leads to a jump in the index. This shows how wall-crossing happens from the perspective of the gravitational index. A similar mechanism is responsible for wall-crossing when studying an arbitrary number of black holes.

    The paper is organized as follows. In section \ref{sec:Einstein_Maxwell} we discuss multicentered Israel-Wilson solutions of pure supergravity in four-dimensional asymptotically flat space,\footnote{The bosonic sector of such theories is just the Einstein-Maxwell theory and therefore the solutions constructed here also contribute to a non-supersymmetric theory with such a sector. 
    Note, however, that generic observables in such non-supersymmetric theories 
     receive contributions from other saddles (not studied in this paper) that do not preserve any supersymmetries. } 
    and revisit the computation of their on-shell action. In section \ref{sec:sugra_with_vectors} we construct multicentered saddles of $\mathcal{N}=2$ supergravity coupled to vector multiplets and show how their on-shell action agrees with microscopic expectations. In section \ref{sec:multicenter_moduli_space}, we initiate the analysis of the finite temperature moduli space of multicentered saddles. Using a mixture of analytic and numerical methods, we analyze in detail the moduli space of a bound state of two black holes and show that the resulting solution space has the topology of $S^3 \times S^1$. We explain how the full moduli space of solutions gets lost when passing through the wall of marginal stability, both for the special case of two black holes and for the more general case of $N$ black holes with a special choice of coplanar charges previously considered in the literature \cite{Manschot:2010qz}. In section \ref{sec:discussion}, we summarize and discuss future directions. In appendix \ref{sec:volume_of_moduli_space} we discuss natural choices for the symplectic measure on finite temperature moduli space and perform a localization computation for the volume, which allows us to comment on wall-crossing in cases of an arbitrary number of black holes.

\section{Einstein-Maxwell theory and pure supergravity}
\label{sec:Einstein_Maxwell}

We begin with a short review of multicentered black holes in pure supergravity in four-dimensional asymptotically flat space, known as Israel-Wilson-Perjes (IWP) solutions  \cite{Israel:1967wq,Hartle:1972ya,Chrusciel:2005ve}.
Here, we focus on the on-shell action for the IWP solutions, which had been previously considered in \cite{Whitt:1984wk,Yuille:1987vw,Dunajski:2006vs} in different cases. 
We emphasize the crucial points that lead to the result with linear dependence on the inverse temperature $\beta$, 
which is a necessary consistency check for the result to match with a microscopic index.

The way the computation organizes itself when working with harmonic functions is what allows us to later perform a similar computation in the general case of $\mathcal{N}=2$ supergravity coupled to an arbitrary number of vector multiplets. 

The IWP solutions studied in this section are also of interest directly from the perspective of $\mathcal{N}=2$ supergravity with vector multiplets. 
This is because for parallel charges of individual black holes \cite{Boruch:2023gfn} 
and a specific choice of asymptotic boundary conditions for the scalar fields, a general multicentered solution reduces to IWP solutions. 
This connection will be explained more in section \ref{sec:special-cases}.

\subsection{Revisiting Israel-Wilson-Perjes solutions}
\label{sec:IWP_solutions}

The IWP solution is the most general supersymmetric solution of Einstein-Maxwell theory in four-dimensional asymptotically flat space \cite{Tod:1983pm}. It is described in terms of two harmonic functions, $\nabla^2 V =\nabla^2  \tilde{V} = 0$ and takes the form 
\begin{align}\label{eq:IWPmetric}
\dd s^2 = \frac{1}{V \tilde{V}} (\dd \tau + \omega_E)^2 + 
V \tilde{V} \dd \xvec^2 ,  
\qquad 
\bast \dd \omega_E = \tilde{V} \dd V - V \dd \tilde{V} . 
\end{align}
Here $\omega_E = \omega_{E,i} \dd x^i$ is a one-form capturing the angular momentum of the solution, and the time coordinate is assumed to be periodic with period $\beta$. The harmonic functions, $V$ and $\tilde{V}$, contain a total of $N_\text{poles} = N_N+N_S$ poles in $\mathbb{R}^3$ base space, 
\be 
V = 1+ \sum_{i=1}^{N_N} 
\frac{Q_i}{|\xvec- \xvec_i|} , \qquad
\tilde{V} = 1 + \sum_{i=1}^{N_S} 
\frac{Q_i}{|\xvec- \txvec_i|} ,
\qquad N_N = N_S \equiv N , \label{eq:VVtIWP} 
\ee
which we refer to as north poles and south poles, respectively.\footnote{In principle, one could consider the more general solutions

\be 
V = 1+ \sum_{i=1}^{N_N} 
\frac{Q_i}{|\xvec- \xvec_i|} , \qquad
\tilde{V} = 1 + \sum_{i=1}^{N_S} 
\frac{\tilde Q_i}{|\xvec- \txvec_i|} ,
\qquad N_N = N_S \equiv N , 
\ee
with $Q_i \neq \tilde Q_i$ subject to the regularity conditions \eqref{eq:regularity_IW}. For a single black hole which has one pole if $V(\xvec)$ and one in $\tilde V(\xvec)$ the regularity condition implies $Q_1 = \tilde Q_1$. More generally, for an arbitrary number of centers the meaning of these solutions is unclear and goes beyond the scope of this paper. 
}
These poles enter the metric as first-order poles, which captures the fact that, in contrast with extremal Majumdar-Papapetrou solutions, the geometry caps off at some finite proper distance in the bulk as one approaches the poles. 
In general, one can consider cases of an arbitrary number of north poles $N_N$ and south poles $N_S$, in which case the geometry will have a multi-TaubNUT asymptotics. In our case, however, we are interested in asymptotically flat solutions, which require the number of north poles and south poles to be equal, $N_N = N_S = N$. 
The north poles and south poles are then connected in pairs via Dirac-Misner strings of the gauge field $\omega$, and are simply coordinate singularities of the metric, provided the following regularity conditions are satisfied: 
\be
Q_i V(\txvec_i) 
= \frac{\beta}{4\pi} , 
\qquad 
Q_i \tilde{V}(\xvec_i) 
= \frac{\beta}{4\pi}  \,, \qquad i=1,\dots,N\,.
\label{eq:regularity_IW}
\ee 
These regularity conditions 
enforce the smoothness of the geometry at the poles. 
The conditions provide a nontrivial constraint on the moduli space of IWP solutions, restricting the allowed positions $(\xvec_i,\txvec_j)$ of the north poles and south poles in the geometry. Each north-south pair of points connected by a Dirac-Misner string can now be viewed as a single supersymmetric rotating black hole, with the string capturing the nontrivial $S^2$ topology between the north pole and the south pole. In particular, for a single black hole $N=1$, one can explicitly rewrite the metric as Kerr-Newman metric with $M=Q$ and nonzero angular momentum $J$. From this perspective, the regularity condition simply ensures that the $S^2$ does not contain any conical singularities near the poles. 

In the case of individual black holes carrying only electric charges\footnote{One can also consider the case of magnetically charged black holes. This has been considered in e.g.~\cite{Dunajski:2006vs}.}, the gauge field of the solution is given by 
\begin{align}
A = \i \Phi (\dd \tau + \omega_E) - 
\i \mathcal{A}_d
, \qquad 
\bast \dd \mathcal{A}_d =   \dd \left(\frac{V - \tilde{V}}{2} \right) , 
\qquad 
\Phi = \frac{1}{2} \left( 
\frac{1}{V} + \frac{1}{\tilde{V}} 
\right) ,
\end{align}
where we have introduced the electric potential $\Phi$.
For the later purpose of evaluating the action, let us also note the explicit expression for the field strength 
\be 
F = \i \dd \Phi \wedge \dd \tau 
+ \i \dd \Phi \wedge \omega 
- \i V \tilde{V} \bast \dd \chi \,, 
\qquad 
\chi = \frac{1}{2} \left( 
\frac{1}{V} - \frac{1}{\tilde{V}} 
\right) 
,
\ee
where $\chi$ denotes the magnetic potential of the solution. 

The angular momentum of the solution can be read off from the asymptotics of the one-form $\omega$ through 
\be    
\omega_E = 2 \i  \epsilon_{ijk} \frac{J^i x^j \dd x^k}{r^3} + O(r^{-3}) ,
\ee
from which one finds 
\be   
\bold{J} = 
\frac{\i}{2} \sum_{i=1}^N Q_i (\xvec_i -\tilde{\xvec}_i) .
\ee
Similarly, by looking at the emblackening factor near asymptotic infinity, we find that the total mass of the system is given by 
\be 
M_{\text{BPS}} = 
\sum_{i=1}^N Q_i
.
\ee

Lastly, let us review how one can compute the horizon area of the individual black holes in the IWP metric. We will follow the presentation of Yuille \cite{Yuille:1987vw}. As we mentioned above, the Dirac-Misner strings connecting north poles to south poles are simply coordinate singularities capturing the black hole horizons in the geometry.\footnote{Naming the string that connects the north and south poles the ''horizon'' is non-standard since we are working in Lorentzian signature and these surfaces are not Killing horizons for certain Killing vectors. Rather, the name is motivated by the fact that they are the surfaces with minimal area homotopic to any surface that encloses the north and south poles.  } For each individual black hole, one can ``blow up" these singularities by going to elliptic coordinates in terms of which the metric is regular and the horizon is properly represented as an $S^2$. This then allows us to derive a simple formula for the finite temperature area of the black hole horizons in terms of the IWP parameters. For any given black hole $(\xvec_i, \txvec_i)$ the elliptic coordinates $(\gamma, \theta , \phi)$ centered on it can be introduced as 
\begin{align}
x = \frac{1}{2}  x_{i \overline{i}}  \sin \theta \cos \phi \sinh \gamma
, \qquad
y = \frac{1}{2}  x_{i \overline{i}}  \sin \theta \sin \phi \sinh \gamma
, \qquad 
z = - \frac{1}{2} x_{i \overline{i}} \cos \theta \cosh \gamma. 
\end{align}
where 
\be 
x_{i \overline{j}} \equiv |\xvec_i - \txvec_j|
\ee
In terms of these coordinates, the horizon is now described by $\gamma = 0$, with parameter $\theta \in [0,\pi]$ going between the north pole and the south pole. 
Rewriting the harmonic functions 
\be  
V(\xvec) = A(\xvec) + 
\frac{Q_i}{r_i} 
, 
\qquad r_i \equiv |\xvec - \xvec_i|
,
\qquad 
\tilde{V}(\xvec) = B(\xvec) + 
\frac{Q_i}{r_{\overline{i}}} 
,\qquad r_{\overline{i}} \equiv |\xvec - \txvec_i| ,
\ee
the full metric can be written as
\begin{align}
\dd s^2 =& \frac{r_i r_{\bar{i}} (\dd \tau + \omega)^2}
{
(Q_i + A(\xvec) r_i)(Q_i + B(\xvec)
r_{\bar{i}})} 
\\ &+ 
(Q_i + A(\xvec) r_i)(Q_i + B(\xvec)r_{\bar{i}}) 
\left( 
\dd \theta^2 + \dd \gamma^2 + 
x_{i \overline{i}}
\frac{x_{i \overline{i}}^2 \sin^2\theta \sinh^2 \gamma}{4 r_i r_{\bar{i}}} \dd \phi^2
\right)
.
\end{align}
The induced metric on the horizon $\gamma=0$, $\tau = \text{const}$, takes now a simple form  
\be 
\dd s^2_{\text{hor}} = \frac{r_i r_{\bar{i}} (\omega_\phi \dd \phi + \omega_\theta \dd \theta)^2}
{(Q_i + A(\xvec) r_i)(Q_i + B(\xvec)r_{\bar{i}})} 
+ 
(Q_i + A(\xvec) r_i)(Q_i + B(\xvec)
r_{\bar{i}}) 
\dd \theta^2  ,
\ee
and leads to the horizon area 
\be 
\text{Area}_{i} = \int_0^{\pi} \dd \theta 
\int_0^{2\pi} \dd \phi \sqrt{\det h} 
=\int_0^{\pi} \dd \theta 
\int_0^{2\pi} \dd \phi \, \frac{x_{i \overline{i}}}{2} |\omega_\phi^{\hor}| \sin \theta 
= 2\pi \, |\omega_\phi^{\hor}| \, \abs{\xvec_i - \tilde{\xvec}_i} ,
\ee
where $\omega_\phi^{\text{hor}}$ denotes the constant value of $\omega_\phi \dd \phi$ on the horizon. This value is, in turn, fixed by the regularity condition to be
\be 
|\omega_\phi^\hor|  = \frac{\beta}{2\pi} , 
\ee
implying that the area of the horizon is simply 
\be 
\text{Area}_{i} = \beta \abs{\xvec_i - \tilde{\xvec}_i} .
\ee
As $\beta\to\infty$, the separation between the poles vanishes as $1/\beta$ and one recovers the area of the extremal black hole.

Having presented the solution and having discussed its regularity, the next step would be to compute its on-shell action. At this point, one might be tempted to use a result by Wald \cite{Wald:1993nt} that shows the on-shell action satisfies the so-called quantum statistical relation as long as the geometry has a Killing vector and a Killing horizon. This certainly holds in the case of a one-center black hole at finite temperature, both for pure supergravity \cite{Iliesiu:2021are} and in the presence of vector multiplets \cite{Boruch:2023gfn}. 

However, this approach does not work for multi-center black holes and therefore we evaluate the on-shell action from first principles in the next subsection. 
Before doing that, it is instructive to elucidate why the multicenter generalization of \cite{Wald:1993nt} does not trivially hold.
The comments that we now make apply to pure supergravity as well as to more general theories of supergravity coupled to an arbitrary number of vector multiplets that we discuss in section~\ref{sec:sugra_with_vectors}. 

It is useful to first analyze the isometries of our solutions. 
The IWP solution has, in general, a single Killing vector given by
$$
\zeta = \partial_\tau\,.
$$
That $\zeta $ is a Killing vector is clear from the fact that metric~\eqref{eq:IWPmetric} is stationary. 
The squared norm of this Killing vector is 
$$
g_{\mu\nu} \zeta^\mu \zeta^\nu \= \frac{1}{V\widetilde{V}} \,.
$$
The right-hand side of this equation only vanishes when either $\xvec=\xvec_i$, which implies that $1/V=0$, or $\xvec=\txvec_i$, which implies that $1/\widetilde{V}=0$. 
These are simply regular points in the geometry, rather than corresponding to a two-surface.\footnote{This should be contrasted with the extremal multi-center Lorentzian solutions where each pole of the harmonic function indeed correspond to an extremal horizon of a black hole; in that case, the surfaces surrounding these poles in fact have a finite proper area and are Killing horizons for the Killing vector $\partial_t$. }  
The event horizon of each black hole is actually topologically an $S^2$ along the Dirac-Misner string joining each $\xvec_i$ (north poles) with $\txvec_i$ (south poles). 
Thus, the only Killing vector in the geometry $\zeta$ fails to degenerate everywhere along the horizon and only does so at the poles. This implies that the event horizons in our geometry are not Killing horizons, something possibly due to the fact that they are complex in Lorentz signature. Therefore, the derivation of the on-shell action in \cite{Wald:1993nt} does not hold.

The argument in the previous paragraph is actually a bit too quick. 
The reason is that in the presence of a single center a new isometry arises given by the Killing vector $\partial_\phi$ (or more generally it is determined by $\omega_i$ at the horizon).\footnote{We are assuming we work at finite temperature since the gravitational path integral at zero temperatures is ill-defined and might lead to the wrong conclusions that BPS black holes have no entropy. 
At zero temperatures and with a single center, this extra Killing vector is enhanced to an $\text{SU}(2)$ isometry.} 
This generates rotations along the angular momentum axis of the black hole. 
Using $\partial_\tau$ and $\partial_\phi$ one can construct a Killing vector that degenerates at the horizon (in the case of pure supergravity, this can be easily seen in the Kerr-Newman presentation of the geometry \cite{Boruch:2023gfn}). 
This is consistent with the fact that the quantum statistical relation does apply in the evaluation of the on-shell action for the single center black hole \cite{Iliesiu:2021are,Boruch:2023gfn}. 
In contrast, when there are multiple centers, each black hole rotates in a different direction. Therefore, although an approximate Killing vector $\partial_\phi$ exists near each center, there is no global Killing vector with this property.

\subsection{On-shell action and the contribution to the index}

We now want to analyze the IWP solutions from the perspective of the four-dimensional gravitational index\footnote{As written, 
this partition function vanishes due to fermion zero-modes in the gravitino originating from the broken supersymmetry. This is fixed by considering instead the helicity supertrace, but this modification does not affect the classical analysis. 
We assume in the following presentation that the fermion zero modes have been absorbed in the calculation of the traces.} 
\be 
\mathcal{I}_{\text{grav}} (\beta,Q) = 
\Tr( (-1)^F e^{-\beta H}) 
.
\ee
Here, $\beta$ denotes the inverse temperature, and we consider a fixed electric charge at infinity. The case of magnetic or dyonic charges is completely analogous. 
As we are working in the canonical ensemble, the variational principle requires adding an additional electromagnetic boundary term to the Einstein-Maxwell action \cite{Hawking:1995ap} ($\kappa^2 = 8\pi G_N$)
\be 
S_{\text{E}} = - \frac{1}{2\kappa^2} \int d^4 x \sqrt{g} 
(R - F_{\mu \nu}F^{\mu \nu} ) - \frac{1}{\kappa^2} \int_{\partial M} d^3 x
\sqrt{h} K 
-\frac{2}{\kappa^2} \int_{\partial M} d^3 x \sqrt{h} n^\mu  F_{\mu \nu} A^\nu
.
\ee

The contribution of the IWP solutions to the gravitational index can now be computed by evaluating their on-shell action 
\be 
\mathcal{I}_{\text{grav}} (\beta,Q) \supset e^{-I_\text{E}(g_{\mu \nu}^{\text{IW}})}
.
\ee
This was originally studied in \cite{Whitt:1984wk} for the case of electrically charged black holes and in \cite{Dunajski:2006vs} for the magnetically charged case. Below we review the key aspects of these computations and highlight some important ingredients missed in the previous discussions\footnote{More precisely, the result derived in \cite{Dunajski:2006vs} does not include the delta function contributions coming from the centers, which leads to the missing entropy contribution in the final action. On the other hand, \cite{Whitt:1984wk} does not include the electromagnetic boundary term required when working in canonical ensemble with fixed electric charges, which leads to the missing term linear in $\beta$.}.


Let us start with the gravitational part of the action. The bulk Einstein-Hilbert part can be simply evaluated using equations of motion $G_{\mu \nu} = T_{\mu \nu}$, where $T_{\mu \nu}$ is the electromagnetic stress tensor. 
Because in four dimensions the electromagnetic stress tensor is traceless, the Einstein equations imply that the Ricci scalar is everywhere vanishing in the geometry, $R=0$, 
and so the bulk gravitational term does not contribute to the on-shell action. 
For the boundary Gibbons-Hawking term, we consider a fixed $r=r_0$ surface at asymptotic infinity. The terms under the integral evaluate to 
\be 
\sqrt{h}K \= r_0^2 \sin \theta \left( 
\frac{\partial_r (V \tilde{V})}{2 V \tilde{V}} 
+ \frac{2}{r_0}
\right).
\ee
Correspondingly, their contribution to the action is given by
\be 
-\frac{1}{\kappa^2} \int_{\partial M} d^3 x \sqrt{h}K \= 
-\frac{\beta}{G_N} r_0
+ \frac{\beta}{2 G_N} \sum_{i=1}^N 
Q_i  . 
\ee
This term is now regulated, in the usual way, by subtracting the contribution of a flat metric with the same boundary metric. We denote the extrinsic curvature of this metric by $K_0$. 
The effect of the regularization procedure is simply a cancellation of the divergent term proportional to $r_0$. The total Gibbons-Hawking contribution to the action is then given by 
\be 
S_{\text{GH}}^{\text{reg}} \=  
-\frac{1}{\kappa^2} \int_{\partial M} \dd^3 x \sqrt{h}(K-K_0)
\=
\frac{\beta}{2 G_N} \sum_{i=1}^N Q_i
\,.
\ee
Now we turn to the contribution of the electromagnetic fields. The bulk term under the integral can be conveniently rewritten as 
\begin{align} 
\sqrt{g} F^2 &\= \frac{1}{2} \partial_i \left( 
\frac{1}{V^2} \partial_i (V \Tilde{V}) 
\right)
- 
\frac{1}{2} \frac{\partial_i ^2 \Tilde{V}}{V} 
- 
\frac{1}{2} \frac{\Tilde{V} \partial_i ^2 V}{V^2} 
\\
&\qquad +
\frac{1}{2} \partial_i \left( 
\frac{1}{\tilde{V}^2} \partial_i (V \Tilde{V}) 
\right)
- 
\frac{1}{2} \frac{\partial_i ^2 V}{\Tilde{V}} 
- 
\frac{1}{2} \frac{V \partial_i ^2 \Tilde{V}}{\Tilde{V}^2} 
\,,
\end{align}
where $\partial_{i}, i=x,y,z$  
denotes the partial derivative with respect to base space $\mathbb{R}^3$ coordinates and we sum over repeated indices. Integrating and using the regularity constraint, the last terms in both lines will go to zero. Importantly, the second terms in both lines give a non-vanishing contribution proportional to $Q_i^2$
\begin{align}
\int \dd ^4 x \sqrt{g} F^2 &\= \frac{\beta}{2} \int_{S^2} r_0^2 \dd \Omega_2 \left( \frac{\partial_r (V \Tilde{V})}{V^2} 
+ \frac{\partial_r (V \Tilde{V})}{\tilde{V}^2}
\right) 
+ 16 \pi^2  \sum_{i=1}^N Q_i^2 \,.
\end{align}
The final contribution of the bulk electromagnetic term to the on-shell action is therefore now given by
\be 
S_{\text{EM}} \=
\frac{1}{2\kappa^2} \int d^4 x \sqrt{g} F^2 \= 
-\frac{\beta}{2 G_N} 
\sum_{i=1}^N Q_i
+ \frac{\pi}{G_N} \sum_{i=1}^N Q_i^2
\,.
\ee
Note that, if the total action only consisted of the above bulk electromagnetic term and the boundary Gibbons-Hawking terms, the terms linear in $\beta$ would cancel and the temperature-independent entropy term would enter the exponential in the index with the wrong sign. This is resolved now by including the electromagnetic boundary term
\be 
S^{\text{bdy}}_{\text{EM}} \=
- \frac{2}{\kappa^2} \int_{\partial M} \dd ^3 x 
\sqrt{h} \, n_\mu F^{\mu \nu} A_\nu 
\,.
\ee
Notice that the above expression is not invariant with respect to large gauge transformations of the field $A_\mu$. 
To properly evaluate it, one needs to work in the regular gauge, in which the holonomy of the gauge field on each horizon is proportional to identity. 
Finding such a gauge in the multicentered case is however nontrivial. 
Therefore, to compute this contribution, we will instead rewrite it as a bulk term. 
For this, we imagine specifying the gauge field in a regular gauge in patches $m_i$ around each black hole. 
For each patch~$m_i$, we can now integrate by parts
\be 
- \frac{1}{\kappa^2} \int_{m_i} \dd^4 x \sqrt{g} F^2 
\=
- \frac{2}{\kappa^2} \int_{\partial m_i} \dd ^3 x 
\sqrt{h} n_\mu F^{\mu \nu} A_\nu
\,,
\ee
where we used equations of motion $\nabla_\mu F^{\mu \nu} =0$. 
After gluing together all of the patches in the geometry, we have that $ \partial \left(\bigcup_i m_i\right) = \partial M$ and so the total boundary term can be rewritten as
\be 
- \frac{2}{\kappa^2} \int_{\partial M} \dd ^3 x 
\sqrt{h} \, n_\mu F^{\mu \nu} A_\nu 
\= - \frac{1}{\kappa^2} \int_{M} \dd^4 x \sqrt{g} F^2 \,,
\label{eq:IW_boundary_term_as_bulk_term}
\ee 
which implies that on-shell, the boundary term given a contribution to the action 
\be 
S^{\text{bdy}}_{\text{EM}} \= -2 S_{\text{EM}} 
\,.
\ee
The effect of this term is then simply a change in the sign of the bulk electromagnetic term. This also restores the electric-magnetic duality of the partition function \cite{Hawking:1995ap}. 
Remember that when working with a fixed magnetic charge, one does not need to include the electromagnetic boundary term in the action. 
However, the Wick rotated gauge fields for the magnetic case are real, whereas for the electric case, they are purely imaginary. 
The combined effect of this is a sign difference in the on-shell electromagnetic bulk terms, which, as we found above, is then exactly compensated by the electromagnetic boundary term.

With this, the total on-shell action then evaluates to 
\begin{align}
S_\text{total} &\= S_{\text{GH}}^{\text{reg}} + 
S_{\text{EM}}^{\text{bdy}} 
+ S_{\text{EM}} 
\= S_{\text{GH}}^{\text{reg}} - S_{\text{EM}} 
\\
&\=  
\frac{\beta}{G_N} \sum_{i=1}^N Q_i
- \frac{\pi}{ G_N} \sum_{i=1}^N Q_i^2
\,,
\end{align}
which can be written as
\be 
S_{\text{total}} \= \beta M_{\text{BPS}} - \sum_{i=1}^N S_{\text{ext},i} \, .
\ee
Note that even though we worked with a finite temperature solution, the finite answer contains only a trivial dependence on inverse temperature~$\beta$. 
This cancellation of the temperature dependence came directly from imposing the smoothness of the underlying geometry. 
Furthermore, the temperature-independent piece is simply the sum of entropies 
of the corresponding zero-temperature black holes. 
Thus we see that 
the multicentered IWP solutions satisfy the new form of attractor mechanism discussed in~\cite{Boruch:2023gfn}. 
This also provides a nontrivial check of the Gibbons-Hawking prescription for multicentered black holes, as the final result is consistent with the 
interpretation of gravitational partition functions as a trace of a putative quantum mechanical system.

\subsection{Moduli space of two or more black holes}
\label{sec:IWP_moduli_space}

Finally, let us make some comments about the moduli space of the IWP solution with two black holes ($N=2$). After fixing rotations and translations, a two black hole configuration is specified in terms of $12-3-3=6$ Cartesian coordinates, subject to the regularity conditions \eqref{eq:regularity_IW}. The regularity conditions are difficult to solve in general, however, for two black holes one can make progress by noticing that the number of Cartesian coordinates which specify a given configuration (before imposing regularity conditions) coincides with the number of distances between four points: $\binom{4}{2}=6$. One can therefore first solve regularity conditions for some of the distances and then solve for Cartesian coordinates describing positions of poles in the configuration. 

The four regularity conditions \eqref{eq:regularity_IW} actually reduce to three  linearly independent conditions, which are given by 
\begin{align}
Q_1 Q_2 \left( 
\frac{1}{|\xvec_1 - \txvec_2|}- 
\frac{1}{|\xvec_2-\txvec_1|}
\right) &\= 0 
\, ,
\\
Q_1 \left( 
1+ \frac{Q_1}{|\xvec_1 - \txvec_1|} 
+ 
\frac{Q_2}{|\xvec_1 - \txvec_2|} 
\right) 
&\= \frac{\beta}{4\pi} 
\,
, 
\\
Q_2 \left( 
1+ \frac{Q_1}{|\xvec_1 - \txvec_2|} 
+ 
\frac{Q_2}{|\xvec_2 - \txvec_2|} 
\right) 
&\= \frac{\beta}{4\pi} 
\,
.
\end{align}
These equations can be solved explicitly for three of the distances in terms of 
one of remaining three,
which we choose here to be $|\xvec_1 - \txvec_2|$:  
\begin{align}
|\txvec_1 - \xvec_2|&\=  |\xvec_1 - \txvec_2| \, ,
\label{iwpxxxx1}\\
|\xvec_1 - \txvec_1| &\= 
\frac{Q_1^2}{\frac{\beta}{4\pi}-Q_1 - \frac{Q_1 Q_2}{|\xvec_1-\txvec_2|}} \label{iwpxxxx2}
\,,
\\
|\xvec_2 - \txvec_2| &\= 
\frac{Q_2^2}{\frac{\beta}{4\pi}-Q_2 - \frac{Q_1 Q_2}{|\xvec_1-\txvec_2|}} \label{iwpxxxx}
\,.
\end{align}
This specifies the moduli space of two IWP black holes at finite temperature. 
It is a six-dimensional solution space, described in terms of three Euler angles specifying global rotations of the entire configuration and the 
three distances $\{|\xvec_1 - \txvec_2|,|\xvec_1-\xvec_2|,|\txvec_1 - \txvec_2|\}$ (the latter two 
not having been fixed by the solution~\eqref{iwpxxxx1}--\eqref{iwpxxxx}).

It is important to note that not all positive distances lead to a valid solution for the Cartesian coordinates, $\xvec_1$, $\txvec_1$, $\xvec_2$, and $\txvec_2$.  
A solution can be guaranteed by imposing an analogue of triangle inequalities for a number of points larger than three, known as the Cayler-Menger condition \cite{blumenthal1970,distance_geometry},  
which imposes that the volume of any polytope whose vertices are the points $\xvec_1$, $\txvec_1$, $\xvec_2$, and $\txvec_2$ is positive. 
Below we give an intuitive discussion of these conditions, postponing a more detailed discussion of the Cayley-Menger condition to section~\ref{sec:2BH_moduli_space}.

In general, the smoothness conditions constrain and bound the moduli space in some directions in the parameter space. 
However, in pure supergravity we find that there are always unbounded directions in the $N$-black hole moduli space\footnote{This should be contrasted with the situation in the presence of vector multiplets we analyze later in section~\ref{sec:sugra_with_vectors}, where the moduli space can be compact.}.
This can be roughly understood as follows. 
The $N$-black hole moduli space is parametrized by~$6N$ Cartesian coordinates, and 
generically has $2N-1$ linearly independent regularity constraints.
After fixing the center-of-mass position of the black holes, this leads to a $4N-2$ dimensional moduli space.
To find examples where the distances between different black holes diverge, we analyze the 
constraint equations and their approximate scaling properties when some of the distances are large. 
For example, we can take $|\xvec_i - \xvec_j| \to \infty$ for $\forall i\neq j$, fixing the distance between $|\xvec_i - \tilde\xvec_i|$ to be that in the respective single black hole case i.e.~$|\xvec_i - \tilde\xvec_i| = \frac{Q_i^2}{\frac{\beta}{4\pi}  - Q_i}$.
In this limit, we see that all smoothness conditions are fully solved.

We note that all the solutions described above are only valid for temperatures that are sufficiently small such that the distances are positive (for example, if $\beta$ is small enough, the denominator of \eqref{iwpxxxx2} and \eqref{iwpxxxx} can become negative). This requirement also appears in supersymmetric rotating Kerr-Newman black holes studied in \cite{Iliesiu:2021are} as well as supergravity in the presence of vector multiplets \cite{Boruch:2023gfn}.

We can contrast the moduli space at finite temperature with the moduli space in pure supergravity for extremal Lorentzian solutions. 
The zero temperature limit of our saddles is the Euclidean version of the Majumdar-Papapetrou solutions, which are the relevant extremal solutions in pure supergravity. 
In this limit, the regularity conditions imply that the  base space distance in the IWP saddles between each north-south pole pair goes to zero; 
therefore, in this limit, the blackening factor in the metric acquires a double pole. 
The location of such double poles determines the location of each black hole horizon at externality. 
In this limit there are no remaining regularity conditions to satisfy and the location of all poles in base space can be arbitrarily chosen. 
Therefore, the moduli space of solutions at extremality is $3N-3$ dimensional (after fixing the location of the center-of-mass of the black holes), in contrast to the $4N-2$ dimensional moduli space discussed above.

Lastly, let us comment on the relevance of the above moduli space for the case of $\mathcal N=2$ supergravity coupled to vector multiplets studied in subsequent sections. In \cite{Boruch:2023gfn} we have shown that the IWP solutions also appear directly in $\mathcal N=2$ supergravity with vectors whenever one studies multicenter configurations with parallel charges and the scalars at infinity are chosen to take the old attractor value. In such cases, the black holes do not form dyonic bound states at extremality. Correspondingly, the number of regularity conditions discussed in \ref{sec:2BH_moduli_space} gets reduced, and the dimension of the solution space is increased. In this way, the IWP solutions provide a special case of multicentered solutions of $\mathcal N=2$ supergravity with vectors subject to the smallest number of constraints.

\section{$\mathcal{N}=2$ supergravity coupled to vectors}
\label{sec:sugra_with_vectors}

We now examine multicentered saddles for the gravitational path integral of $\mathcal{N}=2$ supergravity coupled to $n_V$ vector multiplets. Here, we will briefly introduce the necessary notation and refer the reader to \cite{Boruch:2023gfn,Mohaupt:2000mj} for a more in-depth discussion of the theory. 

The bosonic part of the theory contains fields $(g_{\mu \nu}, A^0)$ of the graviton multiplet together with $n_V$ copies of scalar fields and gauge fields $(t^A, A^A)$, $A=1,\dots,n_V$, which form the bosonic content of a vector multiplet. The action is given by  
\be
\kappa^2
\, 
S \= \frac{1}{16\pi} \int \dd^4 x \sqrt{-g} R -  
\int G_{A\bar{B}} \, \dd t^A \wedge \ast \dd\Bar{t}^{B} 
- \frac{1}{16\pi} \int F^I \wedge G_I \,.
\label{eq:Lorentzian_action}
\ee
The index $I=0,1,\dots,n_V$ on the gauge field strengths includes the ones coming from the vector multiplets ($F^A$) as well the graviphoton field strength ($F^0$). 
The metric~$G_{A\bar{B}}$ on the space of scalar fields $t^A = B^A + \i J^A$ is explicitly given in terms of Kahler potential~$\mathcal{K}$ as
\be 
G_{A\bar{B}} \= 
\frac{\partial^2 \mathcal{K}}{\partial t^A \partial \bar{t}^B} \,,
\qquad \mathcal{K} \= -\log \left(\frac{4}{3} D_{ABC} J^A J^B J^C \right) \, .
\ee
The gauge fields in the action are coupled to the scalars through the scalar-dependent period matrix $-\mathcal{N}_{IJ}$ 
\be 
\label{FGrel}
G_I \= -\Im \, \CN_{IJ}   \ast F^{J} - \Re \, \CN_{IJ}  F^{J} 
\,, \qquad
\CN_{IJ}  \= \overline{F}_{IJ} + \i \frac{N_{IK} X^K N_{JL} X^L}{X^M N_{MN} X^N} .
\ee
Here we expressed the scalar fields in terms of projective coordinates on the scalar manifold $t^A \equiv X^A/X^0$, and introduced a prepotential $F(X)$ whose derivatives we denoted as
\be
F_{IJ} (X) \= \partial_I \partial_J F (X) \,, \qquad N_{IJ} \= F_{IJ} - \overline{F}_{IJ}  \,.
\ee
For the purpose of this paper, we will focus on the simple form of the prepotential given by 
\be 
F \= \frac{1}{6} D_{ABC} \frac{X^A X^B X^C}{X^0} . 
\ee
As there are $n_V+1$ gauge fields in the theory, the charges are now represented as $(2n_V+2)$-component vectors $\Gamma = (\Gamma^0 ,\Gamma^A ,\Gamma_A ,\Gamma_0 )$, with $\Gamma^I$ and $\Gamma_I$ denoting magnetic and electric charges, respectively. 
In this notation, the electric-magnetic duality of the theory acts on the charges as symplectic transformations $\text{Sp}(2n_V+2, \mathbb{R})$, and in our conventions, the corresponding duality-invariant product is given by
\be
\langle A ,  B \rangle 
\=  A^I  B_I -  B^I A_I 
\; \equiv \; I_{\alpha \beta}
A^{\alpha} B^\beta \,. 
\ee
It is also convenient to repackage the information about scalar fields into a so-called normalized period vector 
\be 
\Omega \= \frac{1}{\sqrt{D_{ABC} J^A J^B J^C}}
\left( 
-1 , - t^A , 
- \frac{t_A^2}{2} , \frac{t^3}{6}
\right)
,\qquad t_A^2 \; \equiv \; D_{ABC} t^A t^B , 
\qquad t^3 \equiv D_{ABC} t^A t^B t^C ,
\ee
which transforms under duality transformations as a symplectic vector and satisfies $\langle 
\Omega , \overline{\Omega} 
\rangle = - \i$.

\subsection{Extremal Bates-Denef solutions}
\label{sec:extremal-bates-denef}

A general Lorentzian supersymmetric solution in $\mathcal{N}=2$ supergravity with vectors can be summarized as
\begin{align}
\dd s^2 &\= - \frac{1}{\Sigma(H)}(\dd t + \omega)^2 + \Sigma(H) \dd \xvec^2 , 
\qquad 
\bast \dd \omega \= \langle \dd H , H \rangle ,
\label{eq:extremal_Bates_Denef_metric}
\\
\label{eq:extremal_Bates_Denef_scalar}
t^A (H) &\= \frac{X^A}{X^0} \=
\frac{\i H^A + \partial_{H_A}\Sigma}{\i H^0 + \partial_{H_0}\Sigma}\,, 
\\ 
A^{\alpha} &\= I^{\alpha \beta} \partial_{H^\alpha} \log (\Sigma) (\dd t+ \omega) + \mathcal{A}_d^\alpha \,, 
\qquad 
\dd \mathcal{A}_d^\alpha \= \bast \dd H^\alpha \,.
\label{eq:extremal_Bates_Denef_gauge_field}
\end{align}
The metric is given in terms of a symplectic vector of harmonic functions $H^\alpha$, a homogeneous function of degree two $\Sigma( \lambda H)= \lambda^2 \Sigma(H)$, and a one-form $\omega$ which encodes angular momentum. The function $\Sigma(H)$ is often called the entropy function and can be explicitly determined from the attractor equations after specifying a Calabi-Yau manifold \cite{Shmakova:1996nz}
\be \label{eq:SigmaH}
\Sigma(H) \= \langle H , \Omega_*(H) \rangle 
\langle  H , \overline{\Omega}_*(H) \rangle \, .
\ee
Here, $\Omega_*(H)$ is to be understood as the solution to the generalized attractor equations 
\be \label{genstabOm}
 \i \bigl(\overline{Z}(H;\Omega) \, \Omega - Z(H;\Omega) \, \overline{\Omega} \,\bigr) \= H\,,
\ee
which are imposed on the solution by supersymmetry, and is, therefore, fully determined after choosing the harmonic functions.
The harmonic functions encode information about a given multicentered configuration through positions of their poles $x_{\BH,i}$ in the $\mathbb{R}^3$ base space 
\be 
H(x) \= h + \sum_{i=1}^N \frac{\Gamma_i}{|\xvec - \xvec_{\BH,i}|} 
\,, \qquad
h \= \i (e^{-\i \alpha_\infty} \Omega_\infty - e^{\i \alpha_\infty} \overline{\Omega}_\infty) \,.
\label{eq:extremal_Bates_Denef_harmonic_function}
\ee
Each pole represents now a single extremal black hole, and $\Gamma_i$ denote the charges carried by individual black holes. 
Note that, at this stage, the poles of the harmonic functions enter the metric as second-order poles. 
This captures the information that the near-horizon region of each black hole contains an infinitely long AdS$_2$ throat. 
The constant $h$ depends on asymptotic scalar moduli through $\Omega_\infty \equiv \Omega(t^A_\infty)$ and on the total asymptotic charge $\Gamma= \sum_{i}\Gamma_i$ through the equation~$\i \alpha_\infty = \arg(Z(\Gamma,\Omega_\infty))$, 
which is the requirement of the appropriate fall off of the one-form $\omega$. 
This also ensures that asymptotically $\Sigma(H)|_{r\to \infty} = 1$.

The positions $\xvec_{\BH,i}$ of the black holes in the metric cannot be completely arbitrary. The condition for no closed timelike curves in the geometry imposes an integrability condition for each black hole in the geometry 
\be 
\langle \Gamma_i , H(\xvec_i) \rangle = 0 , 
\qquad i= 1 , \dots , N . 
\label{eq:extremal_Bates_Denef_integrability}
\ee
From the expression for $h$, it can be explicitly verified that $\langle \Gamma , h \rangle = 0$. This then implies that one of the above equations is a linear combinations of the others, as  
\be 
\sum_{i=1}^N \langle \Gamma_i , H(\xvec_i) \rangle = \langle 
\Gamma, h  \rangle = 0 .
\ee
The resulting set of equations consists of $N-1$ linearly independent equations, which act as a constraint on the possible positions of the black holes. After further accounting for the overall translational symmetry of the center of mass, we expect the moduli space to be $3N -3 - (N-1) = 2N-2$ dimensional.

The main case we will investigate in this paper is a bound state of two black holes $N=2$. In this case the distance between two black holes forming the bound state is uniquely fixed from the integrability condition, and the space of solutions simply forms an $S^2$ given by rotations of $\xvec_{\text{BH},1}-\xvec_{\text{BH},2}$. We discuss physical effects captured by this bound state in section \ref{sec:2BH_moduli_space} below. 

Lastly, let us note that even though individual black holes in the extremal bound state are not rotating, a generic multicentered configuration can carry angular momentum carried by the electromagnetic field of the system 
\be 
\mathbf{J} \= \frac{1}{2} \sum_{i=1}^N 
\langle h, \Gamma_i \rangle 
\xvec_{\text{BH},i} 
\,.
\ee
This angular momentum is purely topological, in the sense of being independent of distances between the black holes, as can be seen by using the integrability condition \eqref{eq:extremal_Bates_Denef_integrability}, which allows us to rewrite 
\be 
\mathbf{J} \= \frac{1}{2} \sum_{i,j=1}^N 
\langle \Gamma_i, \Gamma_j \rangle 
\frac{\xvec_{\text{BH},i}-\xvec_{\text{BH},j}}{|\xvec_{\text{BH},i}-\xvec_{\text{BH},j}|}
\, .
\label{eq:extremal_angular_momentum}
\ee
As we will see below, this stops being the case at finite temperature. There, individual black holes start rotating and, correspondingly, the angular momentum starts depending on explicit distances in the system.

\subsection{Multicentered Bates-Denef at finite temperature}
\label{sec:multicentered_Bates_Denef_at_finite_temperature}

We now move on to the main subject of the paper, which are Euclidean multicentered non-extremal supersymmetric solutions. To write down these solutions, we start with a Wick rotated extremal solution described above 
\be 
\dd s^2_{4d} \= \frac{1}{\Sigma(H)} (\dd t_E + \omega_E)^2 + \Sigma(H) \dd x^m\dd x^m \, , 
\qquad 
\bast \dd \omega_E \= \i \langle \dd H , H \rangle ,
\label{eq:4d_attractor_saddle_metric}
\ee
and follow the procedure used previously in a single black hole case \cite{Boruch:2023gfn}. Explicitly, for each black hole in the extremal configuration $(\Gamma_i, \xvec_{\text{BH},i})$ we implement the following procedure: 
\begin{itemize}
    \item We separate a double pole of the metric $\xvec_{\text{BH},i}$ into a north pole $\xvec_i$ with coefficient $\gamma_i$ and a south pole $\txvec_i$ with coefficient $\tgamma_i$. The modified harmonic function now have $N_\text{poles} = 2N$ poles and take the form 
    \be 
    H(x) \= h + \sum_{i=1}^N \frac{\gamma_i}{|\xvec - \xvec_i|} 
    + \sum_{i=1}^N \frac{\tgamma_i}{|\xvec - \txvec_i|} ,
    \label{eq:4d_attractor_saddle_harmonic_function}
    \ee
    and the coefficients are chosen such that each black hole still carries the same total charge as in zero temperature $\gamma_i + \tgamma_i = \Gamma_i$. 
    \item We want to ensure that the geometry caps off smoothly at a finite proper distance in the bulk. This requires the poles of the harmonic functions to enter the metric as first-order poles. Using the generalized attractor equations, one finds that this fixes the coefficients $(\gamma_i,\tgamma_i)$ in terms of the monopole charges $\Gamma_i$
    \begin{align}
    \gamma_i \;\equiv \;\frac{\Gamma_i}{2} + \i \delta_i  
     \= \i \barZ_*(\Gamma_i) \Omega_*(\Gamma_i) \,, 
    \qquad
    \tgamma_i \;\equiv \; \frac{\Gamma_i}{2} - \i \delta_i  
    \= - \i Z_*(\Gamma_i) \barOmega_*(\Gamma_i)
    \,,
    \label{eq:4d_attractor_saddle_charges}
    \end{align}
    where we introduced the notation $\delta_i$ for the dipole charges of each black hole.
    \item Lastly, to ensure smoothness of the resulting geometry, we want to impose that the Dirac-Misner strings appearing in the one-form $\omega_E$ are simply coordinate singularities. This leads to a regularity condition at each of the poles 
    \begin{align}
    \i \langle \gamma_i , H(\xvec_i)  \rangle \= \frac{\beta}{4\pi} 
    , \qquad 
    \i \langle \tgamma_i , H(\txvec_i)  \rangle \= - \frac{\beta}{4\pi}  \,, \qquad     i = 1, \dots , N, \,.
    \label{eq:multicentered_regularity_condition}
    \end{align}
    This condition is exactly analogous to the one appearing in the IWP metric. 
    Note that it is weaker than the integrability equation considered in the extremal case. 
    In Lorentzian signature requiring no closed timelike curves imposes a complete absence of Dirac-Misner strings in the geometry, corresponding to the global existence of one-form~$\omega$. 
    In Euclidean signature, however, the asymptotic time circle is periodic because of our boundary conditions, and a weaker condition, corresponding to the local existence of $\omega_E$, is sufficient. 
\end{itemize}

The configuration carries angular momentum 
\begin{align}
\label{eq:angular-momentum-expression-0}
\mathbf{J} \= \frac{1}{2} \sum_{i=1}^N (\langle h ,\gamma_i \rangle \xvec_i + \langle h ,\tgamma_i \rangle \txvec_i )
\=
\frac{\i}{2} \sum_{i=1}^N \langle h ,\delta_i \rangle (\xvec_{i}-\txvec_{i}) + \frac{1}{4} \sum_{i=1}^N \langle h ,\Gamma_i \rangle (\xvec_{i}+\txvec_{i}) \,.
\end{align}
After using the regularity condition, the angular momentum can be rewritten as
\be 
\label{eq:angular-momentum-expression-1}
\mathbf{J} \= - \frac{\beta}{8\pi \i} \sum_{i=1}^{N} (\xvec_i -\txvec_i) 
+ \frac{1}{2} \sum_{i,j=1}^{N} 
\left[ 
\langle
\gamma_i, \gamma_j
\rangle
\frac{\xvec_i -\xvec_j}{|\xvec_i -\xvec_j|}
+ 
\langle
\gamma_i, \tgamma_j
\rangle
\frac{\xvec_i -\txvec_j}{|\xvec_i -\txvec_j|}
+
\langle
\tgamma_i, \tgamma_j
\rangle
\frac{\txvec_i -\txvec_j}{|\txvec_i -\txvec_j|}
\right]
.
\ee
From this, we see that, in contrast to the extremal case, the angular momentum is not purely topological due to the rotation of each individual black hole.

\subsection{On-shell action and contribution to the index}

To see how the finite temperature saddles contribute to the index, we want to evaluate their Euclidean on-shell actions. This can be done for an arbitrary number of black holes $N$ through a simple generalization of the single black hole computation performed in \cite{Boruch:2023gfn}. Here we will focus on aspects important for the generalization and refer the reader to \cite{Boruch:2023gfn} for some of the details. 

The total on-shell action we need to evaluate is
\be 
-S_{\text{total}} \= -S_{\text{bulk}}  -S_{\text{GHY}}^{\text{boundary}}  -S_{\text{EM}}^{\text{boundary}} 
\,,
\ee
with the Wick rotated bulk term given by
\begin{align}
-S_{\text{bulk}} &\= \frac{1}{16\pi} \int \dd^4 x \sqrt{g} R 
- \int G_{A\bar{B}} \dd t^A \wedge \ast \dd \bar{t}^B 
- \frac{\i}{16\pi} 
\int F^I \wedge G_I
\,,
\\ 
-\i \int F^I \wedge G_I &= 
\int (
\Im \CN_{IJ}\, F^I \wedge \ast F^J 
+ \i\, \Re \CN_{IJ}\, F^I \wedge F^J 
) \,,
\end{align}
where, explicitly, the dual field is 
\be 
G_I \= \i\, \Im \CN_{IJ} \, \ast F^J - \Re \CN_{IJ} \,F^J \,,
\ee
and the Hodge star $\ast$ is understood now as taken in Euclidean signature. The boundary terms consist of the usual regulated Gibbons-Hawking term
\be 
-S_{\text{GHY}}^{\text{boundary}} \=
\frac{1}{8\pi} \int \dd^3 x \sqrt{h} K|_{\text{reg}} 
\,,
\ee
together with an electromagnetic boundary term required to work in an ensemble with fixed electric charges at infinity
\be 
-S_{\text{EM}}^{\text{boundary}} \= 
\frac{\i}{16\pi}
\int \dd^3 x \sqrt{h}\, n_\mu \frac{1}{\sqrt{g}} \epsilon^{\mu \nu\rho \sigma} G_{I, \rho \sigma} A^I_\nu 
\,.
\label{eq:EM_bdry_term}
\ee

\paragraph{Gravity \& scalars:}
We begin with gravity and scalar field contributions to the on-shell action. As in the single black hole case, the Ricci scalar contribution cancels with the scalar fields, leaving only the regulated Gibbons-Hawking term. These evaluates to 
\be 
-S_{\text{GHY}}^{\text{boundary}}
\= 
\frac{1}{8\pi} \int \dd\tau \dd\theta \dd\phi \, r_0^2 \sin\theta \frac{\partial_r \Sigma}{2 \Sigma} 
\= - \frac{\beta}{2} |Z(\Gamma;\Omega_\infty)|
\,.
\ee

\paragraph{Gauge fields:}
The remaining contributions come from the bulk and boundary terms for the gauge fields. For evaluating these terms, it is convenient to introduce electric and magnetic potentials $(\Phi,\chi)$ 
\be 
\Phi^\alpha \;\equiv \; - I^{\alpha \beta} \partial_{\Hv^{\beta}} \log \Sigma \,, 
\qquad 
\dd\chi^\alpha \; \equiv \; \frac{1}{\Sigma} 
(-\Phi^\alpha \langle \dd\Hv, \Hv \rangle +
\dd\Hv^\alpha
) \,,
\label{eq:chemical_potentials_definition}
\ee
in terms of which the electromagnetic fields take the form 
\be 
\mathcal{F} \=  \dd \mathcal{A} \= (F^I,G_I) \= 
\i \dd\Phi \wedge \bigl( \dd t_E + \omE \bigr)
+ \Sigma \bast \dd \chi \,. 
\ee
The bulk part of the on-shell action is now
\begin{align} 
\i \int F^{I} \wedge G_{I} 
\= 
\int \dd^4 x \sqrt{g} 
\left(
\partial_i \Phi^{I} \partial_i \chi_{I} 
+ \partial_i \Phi_{I} \partial_i \chi^{I}
\right) 
\, .
\end{align}
After integration by parts, we find that
\begin{align}
\i \int F^{I} \wedge G_{I}  &\= \int \dd^4 x \, \partial_i (\Phi^I \partial_i H_I +  \Phi_I \partial_i H^I  -  \Phi^I \Phi_I \langle \partial_i \Hv ,\Hv \rangle) 
\\ 
&\phantom{\=} + \int \dd^4 x (
 \Phi^I \Phi_I \langle \partial_i^2 \Hv ,\Hv \rangle 
-\Phi^I \partial_i^2 H_I -  \Phi_I \partial_i^2 H^I ) 
\,.
\label{eq:EM_bulk_two_pieces}
\end{align}
The first line results in a boundary term that evaluates to
\begin{align}
 \int \dd^4 x \; \partial_i (\Phi^I \partial_i H_I +  \Phi_I \partial_i H^I  -  \Phi^I \Phi_I \langle \partial_i \Hv ,\Hv \rangle) &
\= - 4\pi \beta (\phi^I \Gamma_I + \phi_I \Gamma^I ) 
\, ,
\end{align}
where we introduced a notation for chemical potentials at infinity $\phi \equiv \Phi(r\to \infty)$. 
The second line consists now of bulk terms which localize onto contributions coming from the north poles and the south poles 
\be 
 \int \dd^4 x (
 \Phi^I \Phi_I \langle \partial_i^2 H ,H \rangle 
-\Phi^I \partial_i^2 H_I -  \Phi_I \partial_i^2 H^I ) 
\=
- 16\pi^2  
\sum_{i=1}^N
(\delta_{i,I} \Gamma^{I}_i +   \delta^I_i \Gamma_{i,I})
\,.
\ee
To evaluate the above terms, we used the fact that the values of the electric potentials at the poles get fixed by the attractor equations to \cite{Boruch:2023gfn}
\be 
\Phi (\xvec_i) \= \frac{4\pi \i}{\beta} \gamma_i \,, 
\qquad 
\Phi (\txvec_i) \= -\frac{4\pi \i}{\beta} \tgamma_i \,.
\ee
Using this, the total contribution of the bulk electromagnetic term is 
\be 
- \frac{\i}{16\pi}  \int F^I \wedge G_I \= 
\frac{\beta}{4} 
( \phi^I \Gamma_I   + \phi_I \Gamma^I ) 
+ \pi 
\sum_{i=1}^N
(\delta_{i,I} \Gamma^{I}_i +  \delta^I_i \Gamma_{i,I}) \, 
.
\label{eq:action_EM_bulk}
\ee

For the remaining electromagnetic boundary term \eqref{eq:EM_bdry_term}, we use the fact that this term is only sensitive to electric charges. This means in particular that if we rewrite \eqref{eq:EM_bdry_term} as a bulk term, analogously to \eqref{eq:IW_boundary_term_as_bulk_term}, then it will be proportional to \eqref{eq:action_EM_bulk} but with magnetic charges $\Gamma_{i}^I =0$. We find
\be 
-S_{\text{EM}}^{\text{boundary}} 
\= 
\frac{\i}{8\pi} \int F^I \wedge G_I |_{\Gamma_i^I = 0} 
=
-  \frac{\beta}{2} \phi^I \Gamma_I 
- 2 \pi  \sum_{i=1}^N \delta^I_i \Gamma_{i,I} 
\,.
\ee
The effect of the boundary terms is then simply a change of sign in front of terms containing electric charges in \eqref{eq:action_EM_bulk}, which allows us to rewrite the remaining expressions in terms of intersection products  
\begin{align}
- \frac{i}{16\pi}  \int F^I \wedge G_I
-S_{\text{EM}}^{\text{boundary}} 
\= 
\frac{\beta}{4} \langle \Gamma , \phi \rangle 
+ \sum_{i=1}^N
\pi \i  \langle \gamma_i ,\tgamma_i \rangle \,. 
\label{eq:EM_total_part1}
\end{align}
From the definition of electric potentials \eqref{eq:chemical_potentials_definition}, we find that the first term is simply proportional to the asymptotic central charge 
\begin{align}
- \langle  \Gamma , \phi  \rangle
\=
\Gamma^\alpha \partial_{\Hv^\alpha} \Sigma(\Hv)|_\infty \= \langle h , \overline{\Omega}_\infty \rangle 
\langle \Gamma, \Omega_\infty \rangle   
+ 
\langle h , \Omega_\infty \rangle
\langle \Gamma, \overline{\Omega}_\infty \rangle 
\= 2 |Z(\Gamma;\Omega_\infty)| 
\,.
\end{align}

\paragraph{Total on-shell action:}
Altogether, the final expression for the action combines into a simple generalization of the single black hole result 
\be 
-S_{\text{total}} 
\= -\beta \underbrace{|Z(\Gamma;\Omega_\infty)|}_{=M_{\text{BPS}}} \, 
+ \, 
\underbrace{\sum_{i=1}^N \pi \i
\langle \gamma_i , \tgamma_i \rangle}_{\substack{
\= \sum_{i=1}^N S_{\text{ext},i} }} 
\,.
\label{eq:multicentered_action_index_contribution}
\ee
This is one of the key results of our construction. Analogously to IWP solutions and to the single black hole case discussed in \cite{Boruch:2023gfn}, we find that new multicentered saddles satisfy the new form of attractor mechanism: their on-shell action depends on temperature and asymptotic boundary conditions for the scalar fields only through the trivial term involving ground state energy. 
The leftover term is independent of the temperature and factorizes into a sum over extremal entropies of individual black holes. 
This shows, directly from the gravitational path integral, why extremal entropies of multicentered black hole configurations play a role in computing supersymmetric indices without relying on the decoupling of the constituent black holes.

\section{Multi-center moduli space}
\label{sec:multicenter_moduli_space}

Now that we have constructed multicentered saddles for the index and verified how they contribute semiclassically, we will investigate the moduli space of these solutions. Recall, that any multicentered saddle of section \ref{sec:multicentered_Bates_Denef_at_finite_temperature} consisting of $N$ black holes is specified by $N$ monopole charges $\Gamma_{i=1,\dots,N}$, and $N$ pairs of points $(\xvec_i,\txvec_i)_{i=1,\dots ,N}$ in $\mathbb{R}^3$ base space describing the positions of north poles and south poles. Whenever we evaluate the supersymmetric index at fixed charge $\Gamma$ through the gravitational path integral, we are instructed to sum over these saddles; in particular, this means summing over all possible monopole charge splittings $\Gamma = \sum_i \Gamma_i$ as well as integrating over the continuous positions of north and south poles. Note that, because the final on-shell action of any configuration \eqref{eq:multicentered_action_index_contribution} is independent of the positions of the poles, in the exact index computations, one needs to integrate over the continuous space of such solutions.
For example, if in the future one is interested in performing localization for the gravitational path integral   for indices that receive contributions from multicentered saddles (generalizing the results of \cite{Iliesiu:2022kny}), it is therefore critical to understand the moduli space of solutions one should sum over.

From now on, we work with a given charge splitting $\Gamma_{i=1,\dots,N}$, and we ask what is the space of possible values of $2N$ points $(\xvec_i,\txvec_i)_{i=1,\dots ,N}$. 
We assume that the positions are purely real $\xvec_i,\txvec_i \in \mathbb{R}^3$. 
It is entirely not obvious that one should impose such a restriction on the space of solutions, as we are dealing with complex saddles to the gravitational path integral and could imagine working with a bigger moduli space.
Nevertheless, we analyze the moduli space under the assumption that it is always possible to deform the contour of the complex gravitational saddle in such a way
that the positions of the poles end up real. Here, we imagine something similar to what happens with the supersymmetric rotating Kerr-Newman black hole, where after initial Wick rotation 
of Lorentian solution the positions of the poles are purely imaginary at $(-\i a,\i a)$, 
however, it is possible to further Wick rotate the parameter $a \to \i a_E$ to make the positions real. Our assumption can be viewed as demanding that this is possible for multicentered black hole configurations as well. 

Not all positions of the north/south poles are allowed in the $\mathbb{R}^3$ base space. Because we require the geometry to be smooth, the positions are constrained by the smoothness conditions \eqref{eq:multicentered_regularity_condition}. These are in general $2N -1$ linearly independent complex 
equations\footnote{Summing over all regularity conditions together gives $\langle \Gamma,h \rangle=0$ which is trivially satisfied 
due to expression for $h$ given in \eqref{eq:extremal_Bates_Denef_harmonic_function}. This explains why we have $2N-1$ linearly independent equations and not $2N$.}. Under the assumption that all base space distances are real, we can rewrite these conditions as real equations for $i=1,\dots,N$, 
\begin{align}
-\sum_{j=1 \atop j \neq i}^N \frac{\langle \delta_i, \Gamma_j \rangle+\langle \Gamma_i, \delta_j \rangle}
{2|\xvec_i - \xvec_j|} + 
\sum_{j=1}^N \frac{-\langle \delta_i, \Gamma_j \rangle+\langle \Gamma_i, \delta_j \rangle}
{2|\xvec_i - \txvec_j|} 
&\= 
\frac{\beta}{4\pi} + \langle \delta_i , h \rangle
\label{eq:regularity_condition_as_real_equations1}
\,,\\
\sum_{j=1}^N \frac{\langle \delta_i, \Gamma_j \rangle-\langle \Gamma_i, \delta_j \rangle}
{2|\txvec_i - \xvec_j|} + 
\sum_{j=1 \atop j \neq i}^N \frac{\langle \delta_i, \Gamma_j \rangle+\langle \Gamma_i, \delta_j \rangle}
{2|\txvec_i - \txvec_j|} 
&\= 
- \frac{\beta}{4\pi} -\langle \delta_i , h \rangle
\label{eq:regularity_condition_as_real_equations2}
\,,\\
\sum_{j=1 \atop j \neq i}^N \frac{\langle \Gamma_i, \Gamma_j \rangle-4\langle \delta_i, \delta_j \rangle}
{4|\xvec_i - \xvec_j|} + 
\sum_{j=1 \atop j \neq i}^N \frac{\langle \Gamma_i, \Gamma_j \rangle+4\langle \delta_i, \delta_j \rangle}
{4|\xvec_i - \txvec_j|} 
&\= 
- \frac{\langle \Gamma_i , h \rangle}{2}
\label{eq:regularity_condition_as_real_equations3}
\,,\\
\sum_{j=1 \atop j \neq i}^N \frac{\langle \Gamma_i, \Gamma_j \rangle+4\langle \delta_i, \delta_j \rangle}
{4|\txvec_i - \xvec_j|} + 
\sum_{j=1 \atop j \neq i}^N \frac{\langle \Gamma_i, \Gamma_j \rangle-4\langle \delta_i, \delta_j \rangle}
{4|\txvec_i - \txvec_j|} 
&\= 
- \frac{\langle \Gamma_i , h \rangle}{2}
\,,
\label{eq:regularity_condition_as_real_equations4}
\end{align}
where we used the fact that each north pole carries a charge $\gamma_i = \frac{\Gamma_i}{2} +\i \delta_i$ and each south pole carries $\tgamma_i = \frac{\Gamma_i}{2} -\i \delta_i$. This nonlinear set of equations for $(\xvec_i,\txvec_i)$ is rather complicated and highly impractical to solve explicitly in full generality. In some cases, one can make progress by solving the above equations for distances between the points instead of directly solving for coordinates. In that approach, however, even if we determine all the possible solutions for distances to the above regularity conditions, we are still not guaranteed that such a configuration can be consistently embedded in $\mathbb{R}^3$ base space. In other words, one also needs to impose a generalization of triangle-inequalities, which will ensure that the resulting distances describe a valid configuration of points in $\mathbb{R}^3$.

In this section, we mostly focus on analyzing the simplest case of the two black hole saddle, which corresponds to two north poles and two south poles.
In this setting, it turns out one can explicitly solve the above regularity conditions, and it is possible to determine the full space of two black hole solutions. 
This is also sufficient for the purpose of understanding how wall-crossing appears from the perspective of the gravitational path integral. In section \ref{sec:2BH_moduli_space} we analyze the regularity condition for a bound state of two black holes and explicitly solve it in terms of the distances between the poles. 
We then further determine the full moduli space by imposing a Cayley-Menger condition on the resulting space of distances. 
Before proceeding to discuss wall-crossing, we first generalize our discussion of the moduli space to the case of $N$ black holes in section~\ref{sec:moduli-space-Nbh}. 
With this, in section~\ref{sec:wall-crossing_from_GPI}, we analyze the behavior of the finite temperature moduli space during wall-crossing through a mixture of numerical (for $N=2$) and analytical methods (for $N\geq 2$). 
We will confirm that as one crosses the wall of marginal stability, the full moduli space of finite temperature saddles disappears across the wall.

\subsection{Moduli space of two black holes}
\label{sec:2BH_moduli_space}

We are interested in analyzing the finite temperature Euclidean saddle corresponding to a bound state of two extremal black holes with charges $\Gamma_1$ and $\Gamma_2$ at zero temperature. 
At extremality, these solutions are described by the fields given in equations \eqref{eq:extremal_Bates_Denef_metric}, \eqref{eq:extremal_Bates_Denef_scalar}, \eqref{eq:extremal_Bates_Denef_gauge_field}, with the harmonic function \eqref{eq:extremal_Bates_Denef_harmonic_function} consisting of two poles at points $\xvec_{\BH,1}$ and $\xvec_{\BH,2}$. The black holes will form bound states as long as the intersection product of their charges is nonzero, 
\be 
\text{bound state} \quad \Rightarrow \quad 
\langle \Gamma_1 , \Gamma_2 \rangle \neq  0  
\,.
\ee
This happens because the integrability condition, which follows from requiring no CTCs in the geometry, imposes 
\be 
\langle \Gamma_1 , H(\xvec_{\BH,1}) \rangle = 0 
\qquad 
\Rightarrow \qquad 
|\xvec_{\BH,1}-\xvec_{\BH,2}| \= \frac{\langle \Gamma_1 , \Gamma_2 \rangle}{- \langle \Gamma_1 ,h \rangle} 
\; \equiv \; r_{12}^{\text{BatesDenef}}
\,,
\ee
and therefore the only physically sensible Lorentzian geometry is the one in which the two black holes are located at a finite distance from each other, equal exactly to $r_{12}^{\text{BatesDenef}}$. Correspondingly, for two extremal black holes we do not have any remaining free parameters and for fixed charges $\Gamma_1$,$\Gamma_2$ the moduli space is essentially a point, up to overall rotations of the two black hole configuration. 

An important aspect of the above formula is that the bound state distance carries a dependence on asymptotic scalar moduli through the constant $h$ appearing in the denominator. In particular, assuming that $\langle \Gamma_1, \Gamma_2 \rangle >0$, the resulting configuration only makes sense for scalar moduli such that $\langle \Gamma_1 , h \rangle<0$. Starting at some values of scalar moduli for which the distance is positive, we can always tune it in such a way that $\langle \Gamma_1,h \rangle $ approaches zero and then changes sign. What happens then, is that the two black holes go further and further apart from each other until they get infinitely far away at $\langle \Gamma_1,h \rangle =0$. This point in the scalar moduli space is called the wall of marginal stability, and beyond that point, the bound state solutions stop making sense. In Lorentzian signature, this is interpreted as a loss of states from the Hilbert space, which correspondingly leads to a discrete change in the supersymmetric index. This phenomenon has been extensively studied in the literature, see e.g. \cite{Denef:2007vg,Kontsevich:2008fj,Gaiotto:2009hg,Gaiotto:2010be,Gaiotto:2010okc,Dabholkar:2012nd,Hori:2014tda}, and is known as wall-crossing. 

Our goal now is to understand the bound state moduli space and wall-crossing from the perspective of Euclidean saddles of section \ref{sec:multicentered_Bates_Denef_at_finite_temperature}. We therefore consider a solution described by the metric \eqref{eq:4d_attractor_saddle_metric} with harmonic function containing four distinct poles: two north poles $(\xvec_1, \xvec_2)$ and two south poles $(\txvec_1 , \txvec_2)$, carrying charges $(\gamma_1,\gamma_2)$ and $(\tgamma_1,\tgamma_2)$ respectively. The charges are fixed through expression \eqref{eq:4d_attractor_saddle_charges} uniquely in terms of monopole charges $\Gamma_1$ and $\Gamma_2$. 

The solution is described in terms of $12=4\times 3$ parameters corresponding to positions of north poles and south poles. After fixing the translational and rotational symmetry, we end up with $12-6=6$ real parameters required to describe the system. For two black holes, the nice thing that happens is that this configuration can be conveniently described in terms of $\binom{4}{2} = 6$ distances between the poles, which makes solving the regularity conditions \eqref{eq:regularity_condition_as_real_equations1}--\eqref{eq:regularity_condition_as_real_equations4} easier. After finding the 6 distances, one can then solve for the Cartesian coordinates of the corresponding points in $\mathbb{R}^3$. For each set of 6 distances, one finds a discrete number of solutions for the points. 

We will mostly focus on the generic case in which the charges $\Gamma_1$ and $\Gamma_2$ are chosen such that all the intersection products appearing in the regularity conditions are nonvanishing, that is, we assume
\be 
\label{eq:charge_intersections_assumptions}
\langle \Gamma_1 , \Gamma_2 \rangle, \langle \Gamma_1 , \delta_2 \rangle , \langle \Gamma_2 , \delta_1 \rangle , 
\langle \delta_1 , \delta_2 \rangle \neq 0 . 
\ee
This corresponds to the most constrained case, as will become clear below. 

At first sight, we would expect to be able to solve the $4N-2=6$ real regularity conditions directly for the 6 distances. However, it turns out that within the remaining 6 equations, one is nontrivially linearly dependent on the others, leaving us with 5 independent equations. These can then be solved explicitly for five out of six distances in terms of the remaining distance as 
\begin{align}
|\xvec_1 - \txvec_2| &\= 
\frac{B_{1 \twobar}}{d_1 - \frac{B_{12}}{|\xvec_1 - \xvec_2|}} \,,
\qquad 
|\txvec_1 - \xvec_2| \= 
|\xvec_1 - \txvec_2|
\,,
\qquad 
|\txvec_1 - \txvec_2| \= |\xvec_1 - \xvec_2|\,, 
\label{eq:solutions_regularity_conditions1}
\\ 
|\xvec_1 - \txvec_1| &\=  
\frac{A_{1\onebar} B_{1\twobar}}{(B_{1\twobar}c_1-A_{1\twobar}d_1)
+ \frac{A_{1\twobar}B_{12}+A_{12}B_{1\twobar}}{|\xvec_1 - \xvec_2|}}
\,, 
\label{eq:solutions_regularity_conditions2}
\\ 
|\xvec_2 - \txvec_2| &\= 
\frac{A_{2\twobar} B_{1\twobar}}{(B_{1\twobar}c_2-A_{1\twobar}d_1)
+ \frac{A_{1\twobar}B_{12}-A_{12}B_{1\twobar}}{|\xvec_1 - \xvec_2|}}
\,,
\label{eq:solutions_regularity_conditions3}
\end{align}
where we chose the remaining free parameter to be the distance between the north poles $|\xvec_1 - \xvec_2| \equiv x_{12}$, and we introduced the notation 
\begin{align}
B_{1\twobar}&\= 
\frac{\langle \Gamma_1,\Gamma_2 \rangle}{4}+
\langle \delta_1,\delta_2 \rangle 
\,, 
\qquad
B_{12}\=
\frac{\langle \Gamma_1,\Gamma_2 \rangle}{4}-
\langle \delta_1,\delta_2 \rangle  
\,, 
\qquad
A_{1\onebar} = 
\langle \Gamma_1 , \delta_1 \rangle ,
\\
A_{1 \twobar}&\= 
\frac{\langle \Gamma_1,\delta_2 \rangle-\langle \delta_1,\Gamma_2 \rangle}{2} 
\,, 
\qquad
A_{1 2}\= 
\frac{\langle \Gamma_1,\delta_2 \rangle+\langle \delta_1,\Gamma_2 \rangle}{2}
\,, 
\qquad 
A_{2\twobar} \= 
\langle \Gamma_2 , \delta_2 \rangle \,,
\\ 
c_1 &\= 
\frac{\beta}{4\pi} - \langle h , \delta_1 \rangle \,,
\qquad
c_2\= 
\frac{\beta}{4\pi} - \langle h , \delta_2 \rangle  \,, 
\\
d_1  &\= 
\frac{-\langle \Gamma_1 ,h \rangle}{2} 
\,, \qquad 
d_2 \= \frac{-\langle \Gamma_2 ,h \rangle}{2} \= - d_1
\,. 
\label{eq:defd1d2}
\end{align}
We are therefore left with a continuous family of solutions parametrized by the distance $x_{12}$.
Here, we focused on analyzing a generic case, in which $\langle \Gamma_1 , \delta_2 \rangle,\langle \Gamma_2 , \delta_1 \rangle ,
\langle \delta_1 , \delta_2 \rangle
\neq 0 $. 
If some of the above intersection products vanish, one ends up with fewer linearly independent equations, which can lead to more unspecified distances in the solution. If further $\langle \Gamma_1,\Gamma_2 \rangle = 0$, which corresponds to ``the least interacting" case of all charges being parallel, $\Gamma_i = \lambda_i \Gamma$, 
solution space has the same dimensionality as in the IWP solutions of section \ref{sec:IWP_solutions} and correspondingly the number of independent equations gets reduced from five to three. After further fixing asymptotic scalar moduli to the attractor value, the resulting solution space then reduces to that discussed in section \ref{sec:IWP_moduli_space}. For the rest of the section, we continue working with the most constrained case satisfying \eqref{eq:charge_intersections_assumptions} and show the reduction to IWP in section \ref{sec:special-cases}.

We can now try to understand in more detail the moduli space of the solutions satisfying the regularity conditions. In particular, because we are working in terms of the distances between points in $\mathbb{R}^3$ instead of the coordinates of these points, it is important to remember that not all values of $x_{12}$ will yield a set of distances that can be embedded in $\mathbb{R}^3$.  
To verify whether a set of distances for a given value of $x_{12}$ can be interpreted as distances between four points in $\mathbb{R}^3$, we will use the Cayler-Menger condition \cite{blumenthal1970,distance_geometry}. To state the condition, we think of the four points obtained from the regularity condition as a semimetric space\footnote{Semimetric means simply that the mapping $d: X \times X \rightarrow [0,\infty)$ is positive and symmetric, but not subject to triangle inequalities.} 
$X=\{X_a\}_{a=1,\dots,k}=\{ \xvec_i, \txvec_i \}_{i=1,\dots,\frac{k}{2}}$. The Cayley-Menger condition now states that the space $(X,d)$ is isometrically embeddable in $\mathbb{R}^3$ if for all $k=2,3,4$ 
\be 
\label{eq:CM-determinant}
(-1)^k \, \text{CM}(X_1,\dots,X_k) \; \geq \;0 \,. 
\ee
where the \emph{Cayley-Menger determinant~$\text{CM}$} is defined as the determinant of the following distance matrix
\be 
\text{CM}(X_1,\dots,X_k) \=
\det
\begin{pmatrix}
0 & \mathbf{1} \\ 
\mathbf{1} & \Delta_k
\end{pmatrix} , 
\qquad
\Delta_k \= 
\begin{pmatrix}
0 & d_{12}^2 & d_{13}^2 & \dots & d_{1k}^2 \\
d_{12}^2 & 0 & d_{23}^2 & \dots & d_{2k}^2 \\  
d_{13}^2 & d_{23}^2 & 0 & \dots & d_{3k}^2 \\ 
\vdots  &\vdots & \vdots  & \ddots & \vdots \\
d_{1k}^2 & d_{2k}^2 & d_{3k}^2 & \dots  & 0 
\end{pmatrix}
\,.
\ee
For example, written in terms of the distances of our configuration, the distance matrix $\Delta_4$ takes the form
\be 
\Delta_4 \= 
\begin{pmatrix}
0 & x_{1\onebar}^2 & x_{12}^2 & x_{1 \twobar}^2 \\
x_{1\onebar}^2 & 0 & x_{\onebar 2}^2 & x_{\onebar \twobar}^2 \\  
x_{12}^2 & x_{\onebar 2}^2 & 0 & x_{2\twobar}^2 \\ 
x_{1\twobar}^2 & x_{\onebar\twobar}^2 & x_{2\twobar}^2 & 0 
\end{pmatrix}
\,.
\ee
The Cayley-Menger determinant computes the volumes of $(k-1)$-simplices $v_{k-1}$ formed out of points $(X_1, \dots ,X_k)$ 
\be 
\text{Vol}(v_{k-1})^2 \= \frac{(-1)^k}{(k-1)! 2^{k-1}} \text{CM}(X_1,\dots,X_k) \,,
\ee
and therefore, in simple terms, Cayley-Menger condition ensures that the volume of any simplex formed out of points in $X$ computed in terms of distances is real. For our case of four points, the CM determinant reduces to Heron's formula for the volume of the tetrahedron formed out of all six distances
\begin{align}
\sqrt{\frac{\text{CM}(X_1,\dots,X_4)}{288}}
&\= \text{Vol}(\text{tetrahedron})
\\
&\= \frac{\sqrt{(-a+b+c+d)(a-b+c+d)(a+b-c+d)(a+b+c-d)}}{192 u v w}
\,, \nonumber
\end{align}
where we followed the notation from \cite{distance_geometry}
\begin{align}
a&=\sqrt{x Y Z}\,, & X&=(w-U+v)(w+U+v) \,, & x&= (U-v+w)(U-w+v) \,, 
\\ 
b&=\sqrt{y Z X}\,, & Y&=(u-V+w)(u+V+w) \,, & y&=(V-w+u)(V-u+w)
\\ 
c&= \sqrt{z X Y}\,, & Z&=(v-W+u)(v+W+u) \,, & z&=(W-u+v)(W-v+u)
\\
d&= \sqrt{x y z}\,, & \phantom{Z}&\phantom{=,,}  & \phantom{x}&\phantom{=,} \label{eq:d1d2 equations}
\end{align}
and the six distances $(u,v,w,U,V,W)$ are given in terms of the distances between the poles as
\begin{align}
U&= |\xvec_1 - \txvec_1| \equiv x_{1 \onebar} \,, & V&= |\xvec_1 - \xvec_2|
\equiv x_{1 2}
\,, & W&= |\txvec_1 - \xvec_2| 
\equiv x_{\onebar 2}
\,,
\\
u&= |\xvec_2 - \txvec_2|
\equiv x_{2 \twobar}
\,, & v&= |\txvec_1 - \txvec_2|
\equiv x_{\onebar \twobar}
\,, & w&= |\xvec_1 - \txvec_2|
\equiv x_{1 \twobar}
\,.
\end{align}
Here we also introduced a shorthand notation $x_{ij}, x_{i \overline{j}}, x_{\overline{i}\, \overline{j}}$ for the distances.
Similarly, the determinants of any subset of three points will reduce to Heron's formula for area of a triangle. Imposing that the areas are nonnegative corresponds to demanding that the triangle inequalities are satisfied for any three points in the configuration.

\subsubsection*{Numerical analysis}
It turns out that the Cayley-Menger condition is highly constraining for the bound state moduli space. As a guide, we will start with a numerical analysis of the allowed solutions 
and consider an explicit example in the case with a single vector multiplet $n_V = 1$ (we choose $D_{111} = 6$).
To work with a generic setup, we consider a bound state of two black holes with the following monopole charges 
\be 
\Gamma_1 = \{0,1 ,0,6 \}
, 
\qquad 
\Gamma_2 = \{-4, 0 , 9, 4 \} ,
\ee
for which 
\be 
\Sigma^2 (\Gamma_1) = 24 >0, 
\qquad 
\Sigma^2 (\Gamma_2) = 176 >0, 
\qquad 
\Sigma^2 (\Gamma_1+\Gamma_2) = -381 <0  ,
\ee
which implies that a single black hole saddle with the total charge $\Gamma_1+\Gamma_2$ does not exist, however, a bound state of black holes with respective charges $\Gamma_1$ and $\Gamma_2$ could exist for specific values of scalar moduli. Since there are many signs to keep track of in the solution for the distances, we note that the above choice corresponds to the case
\begin{align}
A_{12} &= \frac{\langle \Gamma_1 , \delta_2 \rangle 
+\langle \delta_1 , \Gamma_2 \rangle}{2} <0 ,
&
A_{1 \twobar} &= 
\frac{\langle \Gamma_1 , \delta_2 \rangle 
-\langle \delta_1 , \Gamma_2 \rangle}{2} <0 ,
\\ 
B_{12} &= \frac{\langle \Gamma_1 , \Gamma_2 \rangle}{4}
-  \langle \delta_1 , \delta_2 \rangle <0
,
&
B_{1 \twobar} &= 
\frac{\langle \Gamma_1 , \Gamma_2 \rangle}{4}
+ \langle \delta_1 , \delta_2 \rangle >0
.
\label{eq:bound_state_numerical_sign_assumptions}
\end{align}
Furthermore, for the above charges $\langle \Gamma_1 ,\Gamma_2 \rangle >0$, in order to choose boundary conditions for which the bound state exists we want to be in a region of moduli space for which $\langle \Gamma_1 ,h \rangle < 0$. 
We choose values $(\tau_R,\tau_I) = (-\frac{3}{2}, \frac{4}{10})$. Plotting $\text{CM}(X_1,\dots,X_4)$ for $\beta = 90$ with respect to $x_{12}$ we find the following: 
\be 
\includegraphics[width=0.5\textwidth]{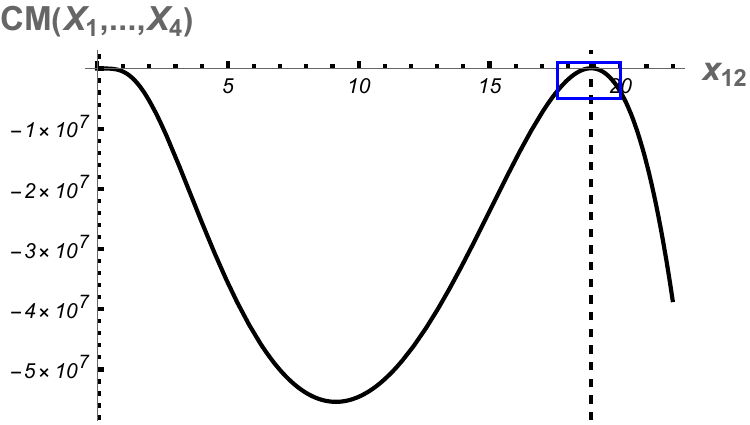}
\nonumber 
\ee
It is clear that for most values of $x_{12}$ the determinant is negative and hence the solutions to the regularity conditions do not represent points in $\mathbb{R}^3$. 
Zooming in closer to the 
dashed line in the figure, 
we find an allowed region as shown in the following plot:
\be 
\includegraphics[width=0.5\textwidth]{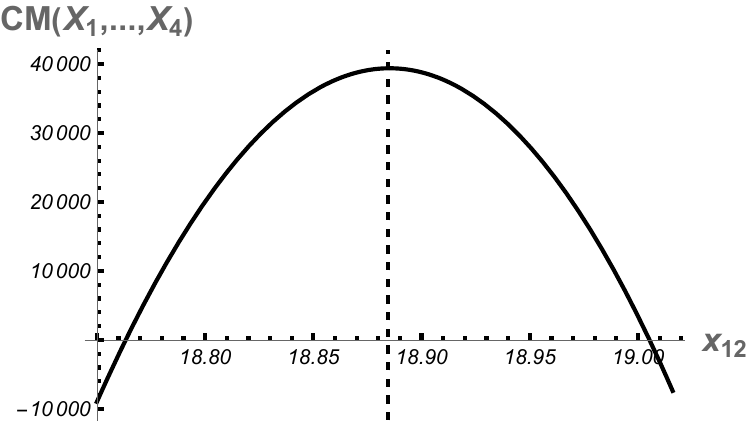}
\nonumber 
\ee
Here we have plotted only the determinant of the largest matrix $\text{CM}(X_1,\dots,X_4)$. However, we also checked that the determinants of all the relevant minors (two of them because of the identifications following from the solutions) are 
all positive in the above region and do not lead to any stronger restrictions on the above moduli space. 

Now we vary the value of~$\beta$ and find that, as we take $\beta$ to be larger and larger, 
the finite temperature moduli space gets more and more concentrated near the dashed line as shown below: 
\be 
\includegraphics[width=0.6\textwidth]{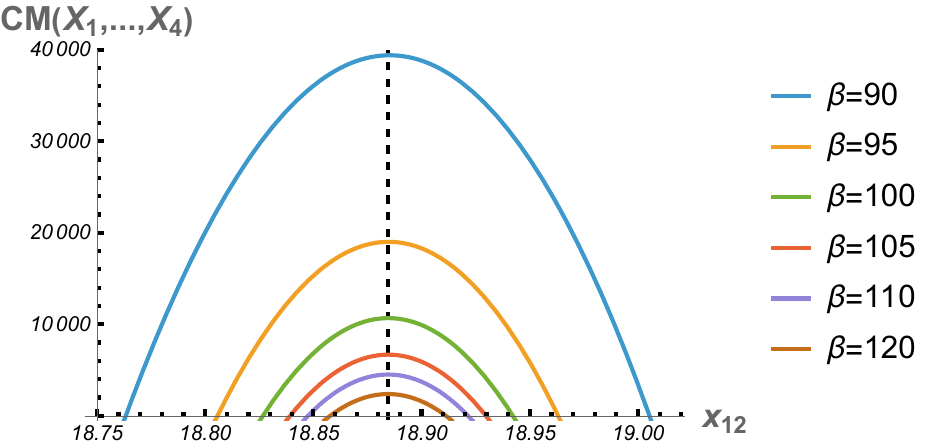} 
\nonumber
\ee
The dashed vertical line in the middle is important. 
It marks the special value of~$x_{12} = x_{12}^*$ where 
the distance between the north poles of the two black holes is exactly equal to the zero temperature bound state distance between the centers of the two black holes, 
\be 
\xgray \;\equiv\; r^{\text{BatesDenef}}_{12} \= - \frac{\langle \Gamma_1 , \Gamma_2 \rangle}{\langle \Gamma_1 , h \rangle}
\,.
\label{eq:x12_gray_line_Bates_Denef}
\ee
As we take $\beta \rightarrow \infty$ the north-south distances of both black holes go to zero $x_{1\onebar},x_{2\twobar} \rightarrow 0$ and in the limit the distance between the north poles becomes the distance between the centers of the black holes $x_{12} \rightarrow 
r^{\text{BatesDenef}}_{12}$. 
What we are seeing is that the zero temperature ($\beta \rightarrow \infty$) bound state solution is ``contained" in the new finite temperature moduli space.

From this numerical analysis, one can expect that the solution given by $x_{12} = \xgray$ will always belong to the finite temperature moduli space (up to the subtleties appearing at high enough temperature, which we discuss more at the end of this section). This turns out to indeed be the case, as we verify analytically in the discussion below. Pictorially, we can view this as the finite temperature moduli space being effectively ``carried around" by the bound state distance $r_{12}^{\text{BatesDenef}}$. This will play an important role in the next section when we analyze wall-crossing.

\begin{figure}[t]
\begin{center}
\includegraphics[width=0.99\textwidth]{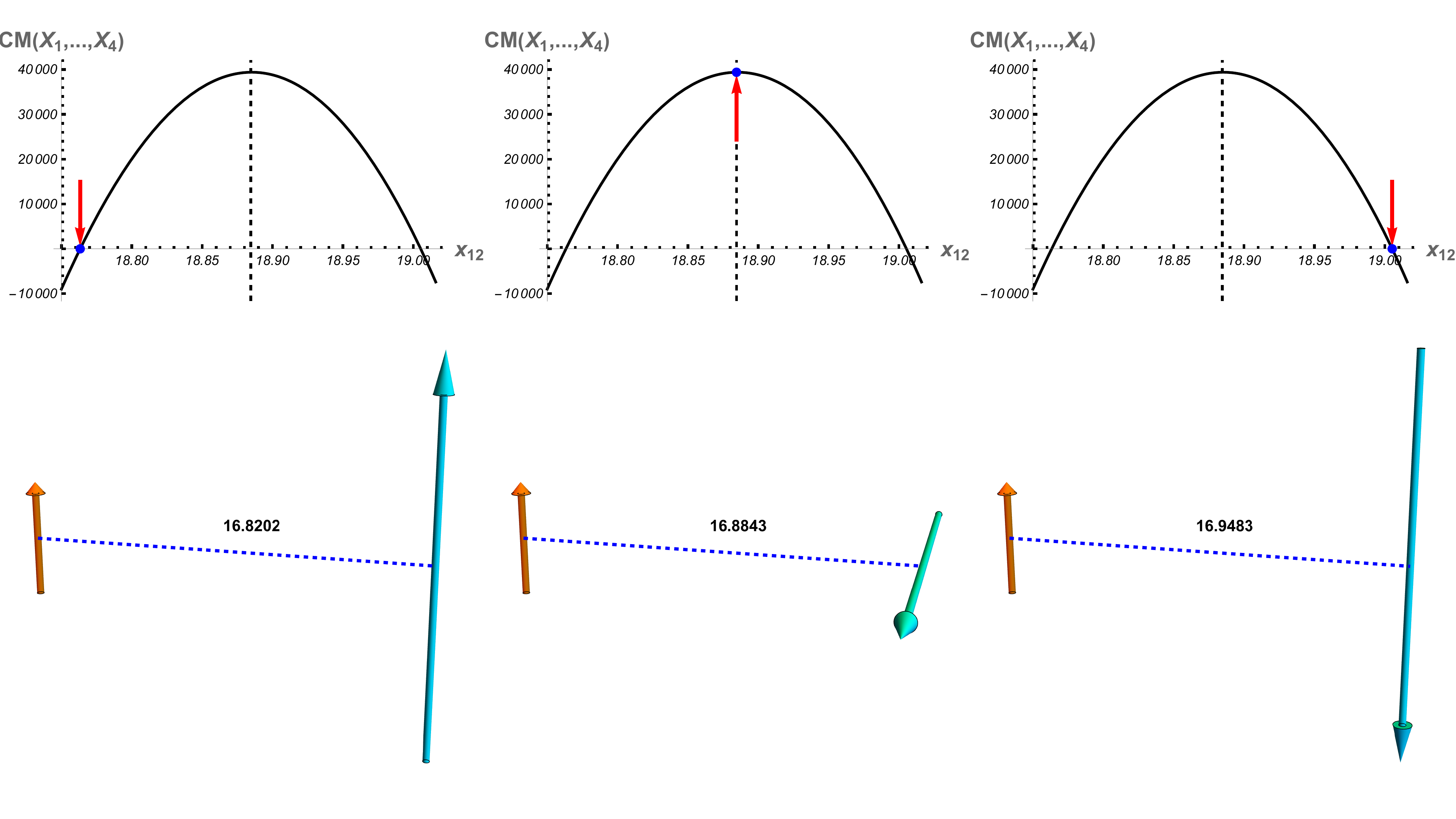}
\caption{The finite temperature moduli space of the bound state saddle. 
The arrows point from the south poles to the north poles of individual black holes. 
The edges of the moduli space correspond to parallel and anti-parallel configurations of the arrows. 
As we move away from the edges, the arrows rotate with respect to each other along parallel planes. 
All the distances captured in the figures vary 
as a function of
$x_{12}$. Importantly, the dashed line represent the values of $x_{12}$ equal to the original zero temperature bound state distance $\xgray = r^{\text{BatesDenef}}_{12}$. This implies that the finite temperature moduli space reduces to the zero temperature moduli space in the extremal limit.}
\label{fig:bound_state_configurations}
\end{center}
\end{figure}

We can also understand more explicitly the configurations represented by our solutions at finite temperature. The examples are presented in figure \ref{fig:bound_state_configurations}. These configurations have been determined by solving for 6 Cartesian coordinates of points, after fixing appropriate symmetries, in terms of the distance solutions for a chosen value of $x_{12}$. Importantly, for each set of distances there are two configurations in $\mathbb{R}^3$ that realize these distances, which are related by a $\mathbb{Z}_2$ reflection. In the plots we focus only on one family of these solutions. 
The three-dimensional arrows in figure~\ref{fig:bound_state_configurations} are pointing from the south pole to the north pole of each individual black hole. The edge cases correspond simply to parallel and antiparallel configurations of the arrows. These represent degenerate (flattened) tetrahedra with vanishing volume. 
One can verify numerically that the distances at the left and right edges, which we denote by~$x_{ij}^{\text{L}}$ and~$x_{ij}^{\text{R}}$, respectively, obey the following equations,
\be 
\label{eq:numerical_edge}
x_{2\twobar}^{\text{L}} \= \frac{(x_{\onebar 2}^{\text{L}})^2-(x_{12}^{\text{L}})^2 }{x_{1\onebar}^{\text{L}}}
\,, 
\qquad 
x_{2\twobar}^{\text{R}} \= 
-\frac{(x_{\onebar 2}^{\text{R}})^2 - 
(x_{1 2}^{\text{R}})^2}{x_{1\onebar}^{\text{R}}}
\, . 
\ee
As we move away from the edges the arrows rotate relative to each other along parallel planes. 
Throughout the moduli space, all the distances captured in the figure slightly change as we vary~$x_{12}$. 
The edges of the moduli space can be viewed as fixed points of $\mathbb{Z}_2$ reflection. 
After reaching any of the edges, two families of solutions coincide and we are able to smoothly switch to reflected solutions. 
In this way, across the full moduli space the blue arrow in the figure traces out a full circle, and the total moduli space of solutions can be topologically viewed as $S^1$. 
We will now confirm this general picture analytically below. 

\subsubsection*{Analytical results}
The lessons derived from numerical analysis can be verified analytically. 
In particular, we can now verify analytically that the configurations plotted in figure~\ref{fig:bound_state_configurations} 
satisfy the regularity constraints and therefore belong to the moduli space.

To see this clearly, let us write down explicitly the Cartesian coordinates of the bound state black holes in terms of the distances in the tetrahedron. 
Aligning the $x$-axis with the first black hole, and the $z$-axis with the vector pointing from the middle point of the first black hole to the middle point of the second black hole, the coordinates of the north and south poles are given by
\begin{align}
\xvec_1 &\= \left\{ \frac{x_{1 \onebar}}{2} ,0 ,0
\right\}
, \qquad
\txvec_1 \= \left\{ - \frac{x_{1 \onebar}}{2} ,0 ,0
\right\}
, \\
\xvec_2 &\= 
\left\{ \frac{x_{ \onebar 2}^2 - x_{12}^2}{2x_{1 \onebar}}
\,,\,- \frac{\eta}{2}
\sqrt{x_{2 \twobar}^2- \frac{(x_{ \onebar 2}^2 - x_{12}^2)^2}{x_{1 \onebar}^2}}
\,,\,\frac{1}{2} 
\sqrt{2(x_{12}^2+x_{\onebar 2}^2) - x_{1 \onebar}^2 - x_{2 \twobar}^2} 
\right\}
\label{eq:Cartesian_coordinates_bound_state_2}
, \quad
\eta \= \pm 1 
\,,  \\
\txvec_2 &\= 
\left\{ -\frac{x_{ \onebar 2}^2 - x_{12}^2}{2x_{1 \onebar}}
\,,\, \frac{\eta}{2}
\sqrt{x_{2 \twobar}^2- \frac{(x_{ \onebar 2}^2 - x_{12}^2)^2}{x_{1 \onebar}^2}}
\,,\,\frac{1}{2} 
\sqrt{2(x_{12}^2+x_{\onebar 2}^2) - x_{1 \onebar}^2 - x_{2 \twobar}^2} 
\right\}, \quad\eta \= \pm 1 
\,, 
\label{eq:Cartesian_coordinates_bound_state_2bar}
\end{align}
where each of the distances $(x_{1 \onebar},x_{2 \twobar}, x_{\onebar 2})$ is a function of $x_{12}$ which we chose to be the moduli space parameter.
Here $\eta$ parametrizes the two families of solutions related by a $\mathbb{Z}_2$ reflection across the $x$-axis. 
These two families of solutions appear because the distance solutions to regularity conditions \eqref{eq:solutions_regularity_conditions1}-\eqref{eq:solutions_regularity_conditions3} can be realized in two ways in terms of the configuration of points in $\mathbb{R}^3$.

The dashed line in the figures sets a special value for the distance~$x_{12}$ as we discussed around~\eqref{eq:x12_gray_line_Bates_Denef}. 
In fact, all the distances take special values on 
dashed line: 
\be 
\xgray 
\= x_{\onebar 2}^{\ast} 
\= x_{1 \twobar}^{\ast} 
\= x_{\onebar \twobar}^{\ast} 
\= -\frac{\langle \Gamma_1 , \Gamma_2 \rangle}{\langle \Gamma_1,h\rangle }
\= r_{12}^{\text{BatesDenef}} \, .
\ee
This corresponds to two possible configurations of points,  
\begin{align}
\xvec_1^{\ast} &\= \left\{ \frac{x_{1 \onebar}^{\ast}}{2} ,0 ,0
\right\}
, \qquad
\txvec_1^{\ast} \= \left\{ - \frac{x_{1 \onebar}^{\ast}}{2} ,0 ,0
\right\}
, \\
\xvec_2^{\ast} &\= 
\left\{ 
0
\,,\, - \eta \, \frac{x_{2 \twobar}^{\ast}}{2}
\,,\, \frac{1}{2} 
\sqrt{4 (x_{12}^{\ast})^2 - (x_{1 \onebar}^{\ast})^2 - (x_{2 \twobar}^{\ast})^2} 
\right\} , \qquad
\eta\= \pm 1 
\,,
\label{eq:gray_line_Cartesian_2}
\\
\txvec_2^{\ast} &\= 
\left\{
0
\,,\, \eta \, \frac{x_{2 \twobar}^{\ast}}{2}
\,,\,\frac{1}{2} 
\sqrt{4(x_{12}^{\ast})^2 - (x_{1 \onebar}^{\ast})^2 - (x_{2 \twobar}^{\ast})^2} 
\right\} , \qquad
\eta\= \pm 1 \,, 
\label{eq:gray_line_Cartesian_2bar}
\end{align}
where the north-south distances take the values 
\be 
x_{1 \onebar}^{\ast} 
\= \frac{\langle \Gamma_1, \delta_1 \rangle}
{\frac{\beta}{4\pi}- \langle h,\delta_1 \rangle
- \frac{\langle \Gamma_1 , h \rangle}{\langle
\Gamma_1, \Gamma_2
\rangle}
\langle \delta_1, \Gamma_2 \rangle
}
\,,
\qquad
x_{2 \twobar}^{\ast}
\=
\frac{\langle \Gamma_2, \delta_2 \rangle}
{\frac{\beta}{4\pi}- \langle h,\delta_2 \rangle
+ \frac{\langle \Gamma_1 , h \rangle}{\langle
\Gamma_1, \Gamma_2
\rangle}
\langle \Gamma_1, \delta_2 \rangle
}
\,.
\ee
It is clear from these equations that the dashed line corresponds to bound state of two black holes rotating in orthogonal directions to each other in $xy$-plane. 
The two solutions given by $\eta = \pm 1 $ are simply related to each other by a reflection across the $x$-axis. 
Importantly, since $\xgray$ is temperature independent, by taking $\beta$ above a certain value one can always make the expression under the square root in \eqref{eq:gray_line_Cartesian_2}, \eqref{eq:gray_line_Cartesian_2bar} positive so that the corresponding coordinates are real. 
Therefore, for large enough $\beta$ the two solutions captured by the dashed line always belong to the moduli space of solutions. 

In the numerical discussion above, we had argued for the fact that, after fixing the overall translational and rotational symmetries, 
we expect the moduli space of finite temperature bound state solutions to be topologically $S^1$. 
We can now  provide additional analytical arguments for 
the same.
The two families of solutions $\eta = \pm 1$ coincide precisely when 
\be 
x_{2 \twobar}^2- \frac{(x_{ \onebar 2}^2 - x_{12}^2)^2}{x_{1 \onebar}^2} \= 0 \,. 
\label{eq:edges_equation_analytic}
\ee
Note that these values also match with our numerical observation for the edges of the moduli space in 
\eqref{eq:numerical_edge}.
The above equation, in general, leads to an eighth-order polynomial in $x_{12}$ with many zeros. 
However, demanding that at $\beta \to \infty$ the solutions collapse to values $x_{12} = r_{12}^{\text{BatesDenef}}$ we find exactly two solutions in perturbation theory, 
\begin{align}
x_{12}^\text{L} &\= \frac{\langle \Gamma_1 , \Gamma_2 \rangle}{-\langle \Gamma_1,h \rangle} 
- \frac{1}{\beta^2} \frac{4\pi^2 \langle \Gamma_1,h \rangle 
\langle \Gamma_1 , \delta_1 \rangle
\langle \Gamma_2 , \delta_2 \rangle
(\langle \Gamma_1 , \Gamma_2 \rangle+ 4 \langle \delta_1 , \delta_2 \rangle)}
{\langle \Gamma_1 , \Gamma_2 \rangle^2}
+ O(1/\beta^3)
\,,
\label{eq:edges_analytic_perturbative_LEFT}
\\
x_{12}^\text{R} &\=
\frac{\langle \Gamma_1 , \Gamma_2 \rangle}{-\langle \Gamma_1,h \rangle}
+ \frac{1}{\beta^2} \frac{4\pi^2 \langle \Gamma_1,h \rangle 
\langle \Gamma_1 , \delta_1 \rangle
\langle \Gamma_2 , \delta_2 \rangle
(\langle \Gamma_1 , \Gamma_2 \rangle+ 4 \langle \delta_1 , \delta_2 \rangle)}
{\langle \Gamma_1 , \Gamma_2 \rangle^2}
+ O(1/\beta^3)
\,,
\label{eq:edges_analytic_perturbative_RIGHT}
\end{align}
which match exactly the numerical edges 
\eqref{eq:numerical_edge}.
Note that here, we assigned the left and right labels under the assumption of \eqref{eq:bound_state_numerical_sign_assumptions}; however, in the opposite case, the labels would simply get flipped. 
Furthermore, we can verify that in the regime of large $\beta$, the last square root in \eqref{eq:Cartesian_coordinates_bound_state_2}, \eqref{eq:Cartesian_coordinates_bound_state_2bar} is always real, i.e., 
\be 
\text{for large }\beta: \qquad 
2(x_{12}^{\text{L/R}})^2+2(x_{\onebar 2}^{\text{L/R}})^2 
- (x_{1 \onebar}^{\text{L/R}})^2 - (x_{2 \twobar}^{\text{L/R}})^2 > 0 \,.
\ee
We, therefore, conjecture that the edges of the finite temperature moduli space are determined through those solutions of \eqref{eq:edges_equation_analytic} which perturbatively behave as \eqref{eq:edges_analytic_perturbative_LEFT},\eqref{eq:edges_analytic_perturbative_RIGHT}. With this, we immediately see that at the edges of the moduli space, two families of reflected solutions, $\eta=\pm 1$, coincide. One can therefore transition from one family of solutions to the other in a smooth fashion, showing that the total space of solutions forms simply an $S^1$. Intuitively, this makes sense, as this is what allows the blue arrow in figure \ref{fig:bound_state_configurations} to complete a full circle as one moves across the space of solutions. Furthermore, we have to account for rotations, which we have previously fixed for the purpose of analyzing the moduli space. Accounting for $SO(3)$ rotations acting at every point in the above solution space, the final form of the bound state moduli space is given by $S^1 \times S^3$. 

\begin{figure}
\begin{center}
\includegraphics[width=0.34\textwidth]{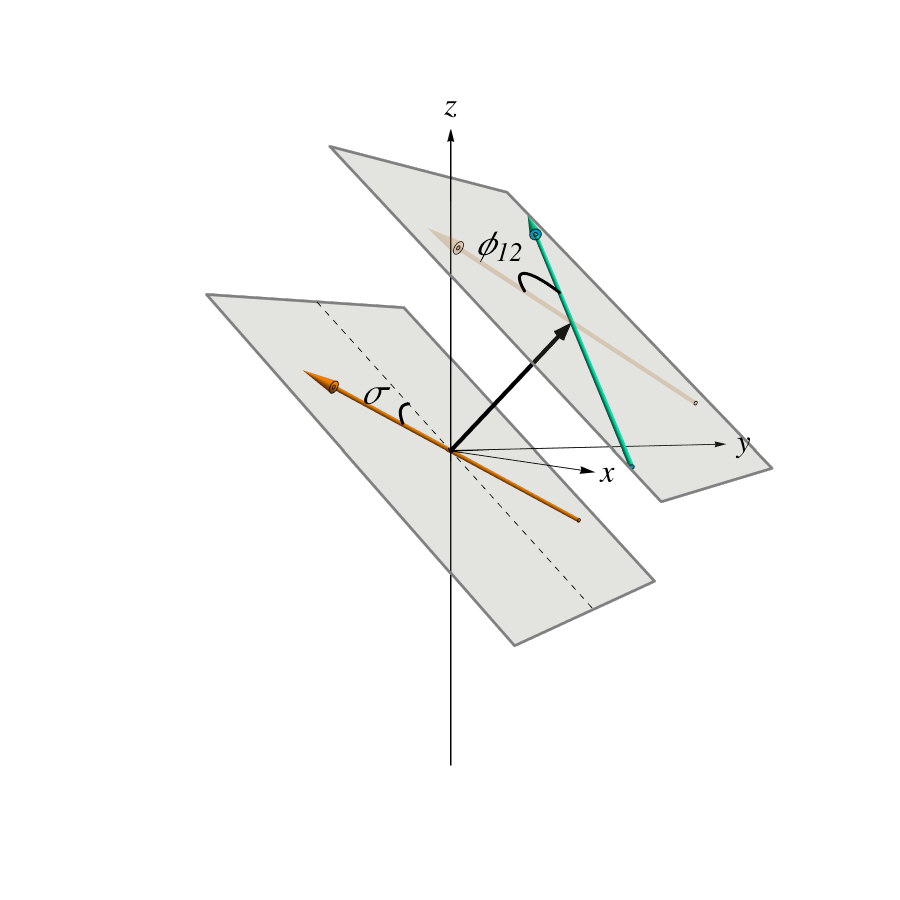}
\caption{
General configuration of a bound state of two black holes in the finite temperature moduli space. The cyan and orange arrows point from the south to the north poles of each black hole and can be viewed as characterizing their individual angular momenta. The black arrow, described in spherical coordinates as $(R_{12},\theta,\phi)$, points between the middle points of the black holes. The regularity condition constrains the black holes to lie in parallel planes to each other. The angle $\sigma$ parametrizes possible directions of rotation of orange black hole (i.e. rotations of orange arrow within its plane), whilst the angle $\phi_{12} \equiv \phi_{12}(R_{12})$ is specified in terms of the distance between the black holes, $R_{12}$, and parametrizes the relative angle between the directions of rotation of orange and cyan black holes. Because the moduli space consists of two copies related by a $\mathbb{Z}_2$ transformation, for each value of $R_{12}$ there are two possible angles: $\phi_{12}$ and $-\phi_{12}$.
}
\label{fig:general_S03_config}
\end{center}
\end{figure}

The $S^1 \times S^3$ moduli space can therefore be parametrized in terms of three Euler angles $(\theta,\phi,\sigma)$ of $SO(3)$ and a relative angle $\phi_{12}(x_{12})$ determined through a point in $x_{12}$ solution space discussed above. We present an example of such a configuration in Figure \ref{fig:general_S03_config}. Here, $(\theta,\phi)$ parametrize spherical angles of the vector $\mathbf{R}_{12}$ connecting middle points of the two black holes and $\sigma$ parametrizes rotations around $\mathbf{R}_{12}$.

\begin{figure}
\begin{center}
\includegraphics[width=0.99\textwidth]{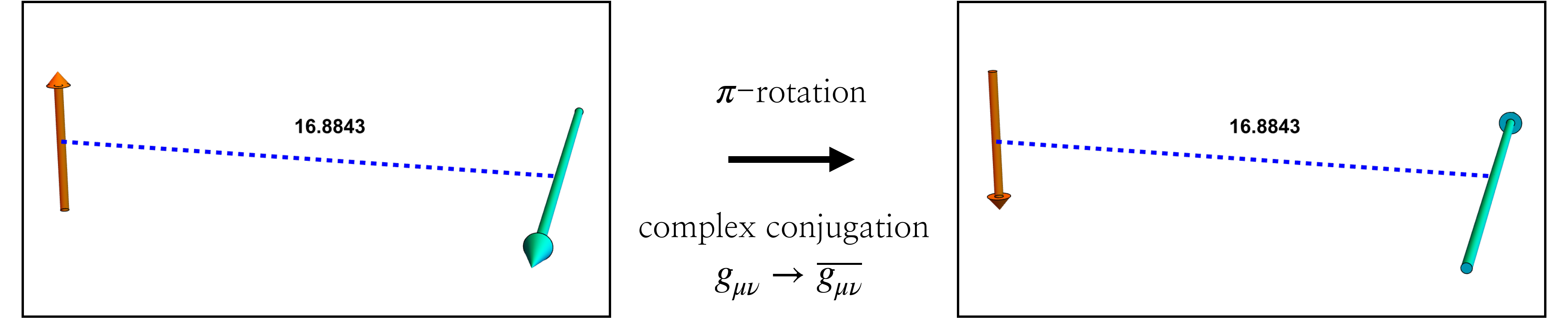}
\caption{The bound states saddles are complex due to the requirement of purely imaginary dipole charges, which implies the existence of saddles related by complex conjugation of the metric. The complex conjugation simply replaces the ``charges" carried by the north pole and the south pole of each of the black holes: $\overline{\gamma_{N/S}}=\gamma_{S/N}$. These saddles can be equivalently obtained through a $\pi$-rotation around the axis connecting the midpoint of the individual black holes. They are therefore already contained in the $S^1 \times S^3$ moduli space discussed in section \ref{sec:2BH_moduli_space}. This is what further leads to the cancellation of imaginary parts of the volume integral obtained from $\omega_{\mathbb{C}}$ discussed in the appendix.}
\label{fig:complex_conjugate_saddles}
\end{center}
\end{figure}

Lastly, let us comment on the complex nature of the saddles we study. Since the solutions we study are complexified, one can wonder whether accounting for the complex conjugate saddles, obtained via $g_{\mu \nu} \to \overline{g_{\mu\nu}}$, enlarges the above moduli space. This turns out to not be the case: complex conjugation acting on the metric exchanges north poles and south poles of individual black holes by sending $\delta_i \to - \delta_i$, which, as we show in figure \ref{fig:complex_conjugate_saddles}, is simply related to the original configuration through rotation by $\pi$ and is therefore already contained in $S^1 \times S^3$ moduli space derived above\footnote{JB thanks Douglas Stanford for a helpful discussion about this point.}.

\subsection{Configurations with non-compact moduli spaces}
\label{sec:special-cases}

The analysis of the moduli space of two black holes at finite temperature done so far assumed that we are sitting at a generic point both in moduli space (scalars at infinity) and that we are working with generic charges. In this subsection, we describe some special features that appear for special choices of these parameters that show qualitatively different features.

If we have two black hole centers with total charge $\Gamma = \Gamma_1 + \Gamma_2$, we can consider a special configuration with a splitting of charges in a parallel fashion
$$
\Gamma_1 \= \Gamma \, \lambda_1,~~~\Gamma_2 \= \Gamma \, \lambda_2,~~~\text{with}~~~\lambda_1 + \lambda_2 \=1\,.
$$
This can only happen when the split is
consistent with charge quantization, and we assume this is the case. 
Let us call $\gamma_N$ and $\gamma_S$ the dipole charges associated to $\Gamma$ as derived via the new attractor equations. In general, the dipole charges of each center are determined by $\gamma_i = \i \bar{Z}_*(\Gamma_i) \Omega_*(\Gamma_i)$ and $\widetilde{\gamma}_i = - \i Z_*(\Gamma_i) \overline{\Omega}_*(\Gamma_i)$ for $i=1,2$. In the present situation $\Gamma_{i} = \lambda_i\,  \Gamma$ and using that under rescaling of the charges $\Omega_*$ is homogeneous of degree 0 and $Z$ of degree 1 we have 
$$
\gamma_i \= \lambda_i \, \gamma_N\,,~~~~ \widetilde{\gamma}_i \= \lambda_i \, \gamma_S\,.
$$
This automatically implies that whenever $\langle \Gamma_1, \Gamma_2\rangle=0$ as above we automatically have $\langle \delta_1, \delta_2 \rangle =0$ as well. 
Note, however, that~$\gamma_N$ and~$\gamma_S$ (and therefore $\gamma_i$ and $\tilde \gamma_i$) are not necessarily parallel. This means that $\langle \Gamma_i, \delta_j \rangle \neq 0$ in general. Indeed, this quantity is proportional to the extremal area of a black hole with charge $\Gamma$. 
Therefore, unless we wish to consider small black holes \cite{Chen:2024gmc}, the above configuration of charges is the best we can assume, and 
we can take the typical situation to be $\langle \Gamma_i, \delta_j\rangle$ not vanishing.

Next, we analyze what happens with the regularity conditions. Consider first \eqref{eq:regularity_condition_as_real_equations3} and \eqref{eq:regularity_condition_as_real_equations4}. From the analysis in the previous paragraph, it is clear that the LHS will vanish identically. This implies that $\langle \Gamma_i , h\rangle =0$ which is equivalent, since $\Gamma_i \propto \Gamma$, to the condition $\langle \Gamma, h \rangle =0$. The net effect of choosing all the centers with parallel charge is then to effectively remove half of the regularity conditions. This also hints that the moduli space will become non-compact, as we explain below.

When selecting the asymptotic moduli, an obvious choice that satisfies the condition $\langle \Gamma,h \rangle =0$ is to pick $h= \Gamma {\sf h}$ with ${\sf h}$ a number. This corresponds precisely to selecting the moduli at infinity to match the single center attractor value for a charge $\Gamma$ as shown in \cite{Boruch:2023gfn}. This is straightforward since far away from the centers, $H \sim h + \Gamma/r\propto \Gamma $, and the stabilization equation implies $\Omega_\infty = \Omega_*(\Gamma)$ takes the old attractor value. This is true not only at infinity, but, in fact, the moduli take the attractor value everywhere in the geometry, even after turning on the temperature or adding multicenter configurations. To see this, decompose $ H=H_N + H_S$, where $H_N$ and $H_S$ involve all the north poles and all the south poles, respectively. 
$$
H_N \= \gamma_N \left( {\sf h} + \frac{\lambda_1}{|\xvec-\xvec_1|} +\frac{\lambda_2}{|\xvec-\xvec_2|} \right)\,,~~~~H_S \= \gamma_S \left( {\sf h} + \frac{\lambda_1}{|\xvec-\txvec_1|} +\frac{\lambda_2}{|\xvec-\txvec_2|} \right).
$$
Equation \eqref{eq:extremal_Bates_Denef_harmonic_function} implies that ${\sf h} = 1/|Z_*(\Gamma)|$ which will be important below. One can then use the new attractor equations to show that 
$$
\i \overline{Z}(H,\Omega_*) \Omega_* \= H_{N}\,,~~~-\i Z(H,\Omega_*) \overline{\Omega}_* \= H_{S}\,.
$$
To derive this, we use that $H_{N/S}$ are proportional to $\gamma_{N/S}$ together with the homogeneity property of the central charge and the fact that $\langle \gamma_N, \Omega_* \rangle = 0$ and $\langle \gamma_S, \overline{\Omega}_* \rangle = 0$. This in turn implies that the old attractor $\Omega = \Omega_*(\Gamma)$ satisfies the stabilization equation for all $\xvec$. This is explained around equation (4.60) of \cite{Boruch:2023gfn} for the case of one black hole, but the derivation generalizes in a straightforward fashion to multicenters with parallel charges so we will not repeat it here. For this choice of charges and moduli, the geometry precisely reduces to the IWP configuration with 
\bea
\Sigma(H)&=& \langle H, \Omega_* \rangle \langle H, \overline{\Omega}_*\rangle \=\langle H_S, \Omega_* \rangle \langle H_N, \overline{\Omega}_*\rangle \nonumber\\
&=& \left(1+ \frac{\lambda_1|Z_*(\Gamma)|}{|\xvec-\xvec_1|}  + \frac{\lambda_2|Z_*(\Gamma)|}{|\xvec-\xvec_2|}\right)\left(1+ \frac{\lambda_1|Z_*(\Gamma)|}{|\xvec-\txvec_1|}  + \frac{\lambda_2|Z_*(\Gamma)|}{|\xvec-\txvec_2|}\right).
\ea
In the second line, we used the fact that $\langle \gamma_S, \Omega_*\rangle = Z(\gamma_S) = Z_* (\Gamma)$ and similarly for $\gamma_N$. This takes precisely the IWP form  \eqref{eq:VVtIWP} where the first factor in the second line is $V$ and the second factor is $\widetilde{V}$ together with identifications 
$$
Q_1 = \lambda_1 \, |Z_*(\Gamma)| ,~~~Q_2=\lambda_2 \, |Z_*(\Gamma)|.
$$
Furthermore, by a simple generalization of the argument in \cite{Boruch:2023gfn} to multiple centers, one can show that the differential equation of $\omega$ also reproduces IWP. The fact that the full solution exactly reproduces IWP implies immediately that the regularity conditions ought to match. More explicitly, one can show that \eqref{eq:regularity_condition_as_real_equations1} and \eqref{eq:regularity_condition_as_real_equations2} reduce to \eqref{eq:regularity_IW}, while we already argued that the remaining equations \eqref{eq:regularity_condition_as_real_equations3} and \eqref{eq:regularity_condition_as_real_equations4} become trivial. To show this, one uses the identities 
$$
\langle \delta, h\rangle = - \i \langle \gamma_N, h\rangle = - \i \langle \gamma_N, \gamma_S \rangle /|Z_*(\Gamma) | = - |Z_*(\Gamma) |
$$
implying that $\langle \delta_i , h \rangle = - Q_i$ and 
$$
\langle \Gamma, \delta \rangle =\i \langle \gamma_N,\gamma_S\rangle = |Z_*(\Gamma)|^2 
$$
implying that $\langle \Gamma_i,\delta_j\rangle = \lambda_i \lambda_j |Z_*(\Gamma)|^2 = Q_i Q_j$. Having done this, the rest is a simple rearranging of \eqref{eq:regularity_condition_as_real_equations1} and \eqref{eq:regularity_condition_as_real_equations2} into \eqref{eq:regularity_IW}. For this reason, the analysis done in section \ref{sec:IWP_moduli_space} on the moduli space of two black holes in pure supergravity also applies in the presence of vector multiplets, as long as the center charges are parallel and the moduli at infinity take the attractor value. For example, the moduli space is non-compact since, as opposed to the generic charges, the two black holes can have an arbitrarily large separation. The ultimate reason behind this is that conditions \eqref{eq:regularity_condition_as_real_equations3} and \eqref{eq:regularity_condition_as_real_equations4} have disappeared in this regime. 

The calculation above can be generalized to multiple centers as long as all charges $\Gamma_i$ are parallel to $\Gamma$, where now $i$ can run over $N$ black holes. All centers will have $\gamma_i$ parallel to $\gamma_N$ and anticenters will have $\widetilde{\gamma}_i$ parallel to $\gamma_S$. If $\Omega_\infty$ is taken to be the solution of the old attractor for $\Gamma$, namely $\Omega_*(\Gamma)$, the scalars will be constant throughout the geometry, and the solution reduces to the IWP for more than two centers as well. The derivation of this fact boils down to the fact that we can still decompose $H=H_N + H_S$ with $H_N$ parallel to $\gamma_N$ and $H_S$ parallel to $\gamma_S$. 

The only remaining special case to consider is when $\Gamma_i$ are parallel to $\Gamma$ but 
the moduli at infinity do not match the old attractor of $\Gamma$, i.e., $\Omega_\infty \neq \Omega_*(\Gamma)$. 
In such a case, the solution is not given by IWP form at all. In particular, due to the new attractor, the scalars cannot possibly be homogeneous in space since their values at each horizon are different from the value at infinity. For example, it is easy to see in equation \eqref{eq:SigmaH} that once the moduli depend on positions, the emblackening factor does not take the IWP form. Nevertheless, the regularity conditions \eqref{eq:regularity_condition_as_real_equations3} and \eqref{eq:regularity_condition_as_real_equations4} are still trivially satisfied. This implies that a configuration where the black holes are far away from each other with $|\xvec_i-\xvec_j|$ large and $|\xvec_i - \txvec_i|$ finite exists, and the moduli space is again non-compact. This can happen even though the geometry is not of IWP type as long as the charges of the centers are parallel. The relevant regularity conditions are very similar to their IWP counterpart but slightly different. In \eqref{eq:regularity_condition_as_real_equations1} and \eqref{eq:regularity_condition_as_real_equations2} we can replace $\langle \Gamma_i,\delta_j\rangle = \lambda_i \lambda_j |Z_*(\Gamma)|^2 $ on the LHS, but the term $\langle \delta_i, h \rangle $ on the RHS does not simplify and depends nontrivially on $\Omega_\infty$. Other than this, the regularity equations take the same form as IWP.

\subsection{Halo saddles}
\label{sec:halo_saddles}

We now make some preliminary observations about a special case with more than two black holes, in which some of the black hole charges are parallel, namely the case of black hole halos. Black hole halos represent the simplest bound state setup one can study for more than two black holes:
here, a single ``core" black hole $(\xvec_c,\txvec_c)$ 
with monopole charge $\Gamma_c$ is surrounded by a halo of $N_h$ black holes 
$(\xvec_i,\txvec_i)$ 
with parallel monopole charges $\Gamma_i = \lambda_i \Gamma_h$, such that $\langle \Gamma_c , \Gamma_h \rangle \neq 0$ and $\sum_i \lambda_i = 1$. From the definition of the dipole charges we therefore also know that $\gamma_i = \lambda_i \gamma_h$ and $\tgamma_i = \lambda_i \tgamma_h$. Since the charges of the black holes in the halo are exactly parallel to each other, they interact with each other in a minimal way, similar to the IWP black holes of section \ref{sec:Einstein_Maxwell}.

The finite temperature regularity conditions in this case consist of $2N_h+2$ complex equations 
\begin{align}
\label{eq:smoothness-finite-T1}
\i \langle \gamma_c , H(\xvec_c)  \rangle  &= \frac{\beta}{4\pi} 
,
\qquad
\i \langle \tgamma_c , H(\txvec_c)  \rangle  = -\frac{\beta}{4\pi} 
, 
\\
\i \langle \gamma_i , H(\xvec_i)  \rangle  &= \frac{\beta}{4\pi} 
,
\qquad
\i \langle \tgamma_i , H(\txvec_i)  \rangle  = -\frac{\beta}{4\pi} .
\label{eq:smoothness-finite-T2}
\end{align}
These equations can be explicitly rewritten as
\begin{align} 
\i \langle \gamma_c ,h \rangle 
+ \frac{\i \langle \gamma_c , \tgamma_c \rangle}{|\xvec_c - \txvec_c|} 
+ \sum_j \lambda_j \frac{\i \langle \gamma_c , \gamma_h \rangle}{|\xvec_c - \xvec_j|} 
+  \sum_{j=1}^{N_h} \lambda_j \frac{\i \langle \gamma_c , \tgamma_h \rangle}{|\xvec_c - \txvec_j|} 
&\= \frac{\beta}{4\pi } ,
\\ 
\i \langle \tgamma_c ,h \rangle 
+ \frac{\i \langle \tgamma_c , \gamma_c \rangle}{|\xvec_c - \txvec_c|} 
+ \sum_{j=1}^{N_h} \lambda_j \frac{\i \langle \tgamma_c , \gamma_h \rangle}{|\txvec_c - \xvec_j|} 
+  \sum_{j=1}^{N_h} \lambda_j \frac{\i \langle \tgamma_c , \tgamma_h \rangle}{|\txvec_c - \txvec_j|} 
&\= -\frac{\beta}{4\pi } ,
\\ 
\forall \, {i=1,\dots, N_h} \qquad
\lambda_i
\left( \i \langle \gamma_h ,h \rangle 
+ \frac{\i \langle \gamma_h , \gamma_c \rangle}{|\xvec_c - \xvec_i|}
+ \frac{\i \langle \gamma_h , \tgamma_c \rangle}{|\txvec_c - \xvec_i|}
+ \sum_{j=1}^{N_h} \lambda_j \frac{\i \langle \gamma_h , \tgamma_h \rangle}{|\xvec_i - \txvec_j|} 
\right)
&\= \frac{\beta}{4\pi } 
\,, 
\\ 
\forall \, {i=1,\dots, N_h} \qquad
\lambda_i
\left( \i \langle \tgamma_h ,h \rangle 
+ \frac{\i \langle \tgamma_h , \gamma_c \rangle}{|\xvec_c - \txvec_i|}
+ \frac{\i \langle \tgamma_h , \tgamma_c \rangle}{|\txvec_c - \txvec_i|}
+ \sum_{j=1}^{N_h} \lambda_j \frac{\i \langle \tgamma_h , \gamma_h \rangle}{|\txvec_i - \xvec_j|} 
\right)
&\= -\frac{\beta}{4\pi } \, .
\end{align}
The $2N_h$ complex equations coming from the halo can now be solved for all the $\{|\xvec_c - \xvec_i|,|\xvec_c - \txvec_i|,|\txvec_c - \xvec_i|,|\txvec_c - \txvec_i|\}$ distances. From the remaining four core equations, three are linearly dependent, and the remaining one fixes the distance $|\xvec_c - \txvec_c|$. The resulting solutions take the form 
\begin{align}
|\xvec_c - \xvec_i| &\= \frac{A_{ch}B_{c\htilde}-A_{c\htilde}B_{ch}}
{d_h A_{c\htilde}+B_{c\htilde}(c_i - A_{h\htilde} \Lambda_i(\txvec_j))}
\qquad i=1,\dots, N_h \,, 
\label{eq:halos_reg_sol1}
\\
|\xvec_c - \txvec_i| &\= 
 \frac{-A_{ch}B_{c\htilde}+A_{c\htilde}B_{ch}}
{d_h A_{ch}+B_{ch}(c_i - A_{h\htilde} \widetilde{\Lambda}_i(\xvec_j))}
\qquad i=1,\dots, N_h \,,  
\\
|\txvec_c - \xvec_i| &\= 
 \frac{-A_{ch}B_{c\htilde}+A_{c\htilde}B_{ch}}
{d_h A_{ch}+B_{ch}(c_i - A_{h\htilde} \Lambda_i(\txvec_j))}
\qquad i=1,\dots, N_h \,,  
\\
|\txvec_c - \txvec_i| &\= 
\frac{A_{ch}B_{c\htilde}-A_{c\htilde}B_{ch}}
{d_h A_{c\htilde}+B_{c\htilde}(c_i - A_{h\htilde} \widetilde{\Lambda}_{i}(\xvec_j))}
\qquad i=1,\dots, N_h \,,  
\\
|\xvec_c - \txvec_c| &\= 
\frac{A_{c\tilde{c}}[A_{c\htilde}B_{ch}-A_{ch}B_{c\htilde}]}
{c_c(A_{c\htilde}B_{ch}-A_{ch}B_{c\htilde}) 
-2 A_{ch}A_{c\htilde} d_h 
+(\Lambda A_{h\htilde}-c_N) (A_{ch}B_{c\htilde}+A_{c\htilde}B_{ch})
}
\,, \label{eq:halos_reg_sol5}
\end{align}
where we introduced the notation
\begin{align}
B_{c\htilde} &= 
\frac{\langle \Gamma_c, \Gamma_h \rangle
+4 \langle \delta_c, \delta_h \rangle 
}{4} , 
\qquad 
B_{ch} = 
\frac{\langle \Gamma_c, \Gamma_h \rangle
-4 \langle \delta_c, \delta_h \rangle 
}{4} ,
\\ 
A_{c\htilde} &= 
\frac{\langle \Gamma_c, \delta_h \rangle
-\langle \delta_c, \Gamma_h \rangle
}{2}
,
\qquad
A_{ch} = 
\frac{\langle \Gamma_c, \delta_h \rangle
+\langle \delta_c, \Gamma_h \rangle
}{2} , 
\\ 
d_c &= - \frac{1}{2} \langle \Gamma_c , h \rangle
, 
\qquad 
d_h = - \frac{1}{2} \langle \Gamma_h , h \rangle
,  
\\ 
c_c &= \frac{\beta}{4\pi} + \langle \delta_c ,h \rangle
,
\qquad
c_i = 
\frac{\beta}{4\pi \lambda_i} + \langle \delta_h ,h \rangle
,
\qquad 
c_N = \frac{\beta N_h}{4\pi} + \langle \delta_h ,h \rangle,
\\ 
A_{c \tilde{c}} &= \langle \Gamma_c , \delta_c \rangle 
, 
\qquad 
A_{h \htilde} = 
\langle \Gamma_h , \delta_h \rangle 
, 
\\ 
\Lambda_i (\txvec_j) &= 
\sum_{j=1}^{N_h} \frac{\lambda_j}{|\xvec_i-\txvec_j|} 
, 
\qquad 
\tilde{\Lambda}_i (\xvec_j) = 
\sum_{j=1}^{N_h} \frac{\lambda_j}{|\txvec_i-\xvec_j|} , 
\qquad 
\Lambda = \sum_{i,j=1}^{N_h} 
\frac{\lambda_i \lambda_j}{|\xvec_i-\txvec_j|} 
.
\end{align}
The total system of $N_h$ halo black holes and a single core is described by $6(N_h+1)$ coordinates: $3(N_h+1)$ coordinates describing north poles, and $3(N_h+1)$ coordinates describing south poles. Removing translations and $SO(3)$ rotations leaves us with $6N_h$ coordinates. 
In total, there is $\binom{2(N_h+1)}{2}$ distances describing the system, so fixing all of them would overconstrain the configuration. 
We can, however, choose a $6N_h$ subset of these distances as the coordinates of the moduli space. 
The regularity conditions described above now fix $4N_h+1$ of those distances, leaving us with $2N_h-1$ dimensional moduli space of the halo saddles. 
It would be interesting to understand the space of possible configurations represented by above equations in more detail. We leave this for future work.

\subsection{Moduli space of $N>2$ black holes}
\label{sec:moduli-space-Nbh}

While in section \ref{sec:halo_saddles}, we have managed to explicitly describe the parametrization of the moduli space of index saddles for $N>2$ black holes in the special case where some of the charges are parallel, in general, it is difficult to solve the regularity constraints \eqref{eq:regularity_condition_as_real_equations1} -- \eqref{eq:regularity_condition_as_real_equations4} to find an explicit parametrization of the moduli space. Nevertheless, below we will make some general remarks about the moduli space $\mathcal M$ for arbitrary $N$ and for generic non-parallel charges.  

Points in $\mathcal M$ need not only satisfy the regularity conditions but also satisfy the positivity of the Cayley-Menger determinant \eqref{eq:CM-determinant} as in the simpler case $N=2$. 
Specifically, given the distance among any choice of 4 points in base space from $X \equiv \{\xvec_i, \tilde \xvec_i\}$, 
the determinant  \eqref{eq:CM-determinant} needs to be semi-positive in order to guarantee that the points $\xvec_i$ and $\tilde \xvec_i$ are embeddable in~$\mathbb R^3$. 
Thus, as for $N=2$, the boundary of $\mathcal M$ is determined by any one of the Cayley-Menger determinants vanishing. 
Generically, only one of the Cayley-Menger determinants vanishes on $\partial \mathcal M$. 
One can consequently associate a locus to the set of boundary points where a specific Cayley-Menger determinant $\text{CM}(X_1, X_2, X_3, X_4)$ vanishes. At the intersection of these loci,  multiple Cayley-Menger determinants vanish, and more than one 4-point subset of $X$ is co-planar; at such points, we expect $\partial \mathcal M$ to be singular. The most singular case is when all points in $X$ become coplanar. As we shall discover shortly, such configurations turn out to be isolated points in $X$ (up to overall rotations) and are useful in studying its properties.  

For example, they are useful in finding the dimension of $\mathcal M$. While to determine this we could, in principle, check the number of independent equations among the smoothness conditions, this is challenging to check at an arbitrary point in moduli. Rather, it is more convenient to find a parametrization of $\mathcal M$ in the neighborhood of one of the coplanar configurations.  This can be obtained by solving \eqref{eq:regularity_condition_as_real_equations1} -- \eqref{eq:regularity_condition_as_real_equations4} to linear order:
 \begin{align}
-\sum_{j=1 \atop j \neq i}^N \frac{\langle \delta_i, \Gamma_j \rangle+\langle \Gamma_i, \delta_j \rangle}
{2|\xvec_i - \xvec_j|^{3}} (\xvec_i - \xvec_j)(d\xvec_i - d\xvec_j) + 
\sum_{j=1}^N \frac{-\langle \delta_i, \Gamma_j \rangle+\langle \Gamma_i, \delta_j \rangle}
{2|\xvec_i - \txvec_j|^{3}}  (\xvec_i - \txvec_j)(d\xvec_i - d\txvec_j)
&\= 
0
\,,\nonumber\\
\sum_{j=1}^N \frac{\langle \delta_i, \Gamma_j \rangle-\langle \Gamma_i, \delta_j \rangle}
{2|\txvec_i - \xvec_j|^{3}} (\txvec_i - \xvec_j)(d\txvec_i - d\xvec_j) + 
\sum_{j=1 \atop j \neq i}^N \frac{\langle \delta_i, \Gamma_j \rangle+\langle \Gamma_i, \delta_j \rangle}
{2|\txvec_i - \txvec_j|^{3}} (\txvec_i - \txvec_j)(d\txvec_i - d\txvec_j) 
&\= 
0
\,,\nonumber\\
\sum_{j=1 \atop j \neq i}^N \frac{\langle \Gamma_i, \Gamma_j \rangle-4\langle \delta_i, \delta_j \rangle}
{4|\xvec_i - \xvec_j|^{3}} (\xvec_i - \xvec_j)(d\xvec_i - d\xvec_j)  + 
\sum_{j=1}^N \frac{\langle \Gamma_i, \Gamma_j \rangle+4\langle \delta_i, \delta_j \rangle}
{4|\xvec_i - \txvec_j|^{3}} (\xvec_i - \txvec_j)(d\xvec_i - d\txvec_j) 
&\= 
0
\nonumber
\,,\\
\sum_{j=1}^N \frac{\langle \Gamma_i, \Gamma_j \rangle+4\langle \delta_i, \delta_j \rangle}
{4|\txvec_i - \xvec_j|^{3}} (\txvec_i - \xvec_j)(d\txvec_i - d\xvec_j)  + 
\sum_{j=1 \atop j \neq i}^N \frac{\langle \Gamma_i, \Gamma_j \rangle-4\langle \delta_i, \delta_j \rangle}
{4|\txvec_i - \txvec_j|^{3}} (\txvec_i - \txvec_j)(d\txvec_i - d\txvec_j) 
&\= 
0
\,,
\label{eq:regularity_condition_as_real_equations4p}
\end{align}
where $\xvec_{i, j}$ and $\txvec_{i, j}$ are evaluated on the coplanar configurations discussed above. These equation are solved by making $d y_i\neq 0$ and $d\tilde y_i \neq 0$ (subject to the constraint that $\sum_i (\dd y_i + \dd \tilde y_i) = 0 $) since the fixed point configuration is constrained to the $x$-$z$ plane. We can also take $\dd x_i = -z_i(p) \dd\epsilon  $,  $\dd\tilde x_i = -\tilde z_i(p) \dd\epsilon  $, $\dd z_i =  x_i \dd\epsilon  $, and $\dd\tilde z_i = \tilde x_i \dd\epsilon  $ for some infinitesimal $d\epsilon$. 

In total, there are thus \be 
\text{dim} \,\mathcal M \= (N_\text{poles}-1) + 1 \= N_\text{poles} \= 2N
 \ee
 coordinates that parameterize the moduli space.\footnote{Several points are noteworthy. First, the fact that there are $2N_\text{poles}-3$ real linearly independent equations among the smoothness conditions is not immediately apparent from the $2N_\text{poles} = 4N$ real equations  \eqref{eq:regularity_condition_as_real_equations1} -- \eqref{eq:regularity_condition_as_real_equations4}. Second, one can associate a non-degenerate symplectic form $\omega_\beta$ (see \eqref{eq:symplectic-form-NBH}) to the moduli space for which the angular momentum $J_z$ is the moment map to the moduli space. This, of course, implies that the moduli space is a symplectic manifold even for $N>2$, though we leave the analysis of the symplectic volume of this manifold to future work.  }

Finally, to see that the coplanar configurations are isolated points in $\mathcal M$ (up to overall rotations), we can simply compare the number of linearly-independent regularity conditions to the number of coordinates required to parameterize such coplanar configurations. For such a count, we can gauge fix rotations and set all points in $X$ in the $x$-$z$ plane. After also gauge-fixing translations (i.e., imposing that $\sum_A \mathbf x_A = 0$), the co-planar solutions are described in terms of $2N_\text{poles} - 3$ coordinates. However, we need to impose the smoothness equations \eqref{eq:regularity_condition_as_real_equations1} -- \eqref{eq:regularity_condition_as_real_equations4}, which, as we discovered above, consist of $2N_\text{poles}-3$ linearly independent equations. Since the number of equations matches the number of undetermined coordinates of the coplanar solutions, we can only have a discrete set of coplanar solutions in the $x$-$z$ plane for any number of poles $N_\text{poles}$. 

\subsection{Detecting when wall-crossing occurs}
\label{sec:wall-crossing_from_GPI}

\textbf{$N=2$: }Since in the case of two-black holes, we have explicitly parametrized the moduli space in terms of the distance between one of the poles of the two black holes, we will first discuss wall-crossing for this simplest case. First, let us review how wall crossing occurs in Lorentzian signature, at zero temperature. Following section \ref{sec:extremal-bates-denef}, the base-space distance between the two black holes is given by the smoothness condition $\<\Gamma_i, H(\xvec_i)\> = 0$, which consequently implies 
\be 
r_{12}^{\text{BatesDenef}} \;\equiv\; |\xvec_{BH, 1} - \xvec_{BH, 2}| \= -\frac{\<\Gamma_1, \Gamma_2\>}{\<\Gamma_1, h\>}\,.
\ee
Since the LHS is manifestly positive, we need $\<\Gamma_1, h\> \geq 0 $ if $\<\Gamma_1, \Gamma_2\> < 0$ and $\<\Gamma_1, h\> \leq 0 $ if $\<\Gamma_1, \Gamma_2\> > 0$ in order for the solution to exist. In both cases, the wall of marginal stability on which wall crossing occurs is given by $\< \Gamma_1, h\> = \<\Gamma_2, h\> = 0$, at which point the base-space distance between the two extremal black holes goes to infinity.

\begin{figure}
\begin{center}
\includegraphics[width=0.7\textwidth]{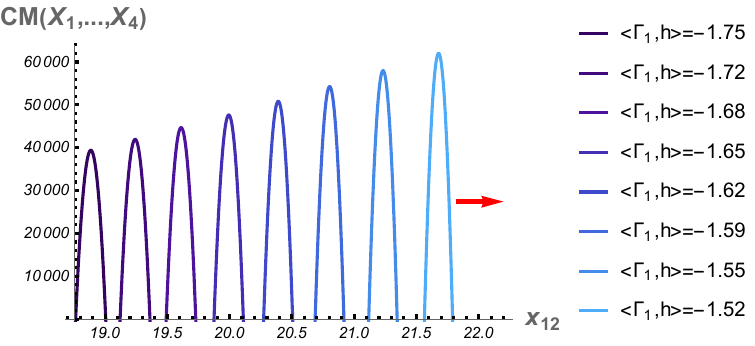}
\caption{The plot of the finite temperature moduli space being effectively ``carried around" by the zero temperature moduli space. Different colors in the plot represent values of $\text{CM}(X_1,\dots,X_4)$ with respect to $x_{12}$ for different choices of $\langle \Gamma_1,h \rangle$. As we slowly increase the values of scalar moduli towards the wall of marginal stability at $\langle \Gamma_1,h \rangle =0$, the allowed configurations move to larger values of $x_{12}$. At the wall of marginal stability, the allowed solutions jump to negative values through infinity. In this way, after passing through the wall, we lose the full finite temperature moduli space of bound states saddles.}
\label{fig:wall_crossing}
\end{center}
\end{figure}

With this in mind, we can now analyze when wall crossing occurs using the finite temperature saddle contributions to the GPI. We will again detect wall crossing by finding the value of $h$ at which saddles with real $\xvec_i$ and $\tilde \xvec_i$ stop contributing to the GPI.  The crucial point follows from expressions for the edges of the moduli space $x_{12}^{\text{L/R}}$ given in \eqref{eq:edges_analytic_perturbative_LEFT} and \eqref{eq:edges_analytic_perturbative_RIGHT}. As one approaches the wall of marginal stability by decreasing $\langle \Gamma_1, h \rangle \to 0$, the whole moduli space moves to larger and larger values of $x_{12}$ as $\sim -1/\langle \Gamma_1, h \rangle$. In other words, the finite temperature moduli space is effectively centered around 
$\xgray$
. 
After passing through the wall at $\langle \Gamma_1,h \rangle =0$, 
$\xgray$
will change its sign, and correspondingly the allowed region of solutions jumps to negative values of $x_{12}$, passing through infinity. In this way, after passing through the wall of marginal stability, we lose the full moduli space of finite temperature saddles, leading to the expected jump in the index. We plot the example of this behavior with respect in Figure \ref{fig:wall_crossing}. Thus, even though the moduli space for the newly found saddles is completely different than the moduli space of Lorentzian solutions even for two black holes, we find that the condition that determines the wall of marginal stability, $\< \Gamma_1, h\> = \<\Gamma_2, h\> = 0$, is the same in both cases.

\noindent \textbf{$N > 2$ in a two-dimensional charge sublattice: } For an arbitrary number of black holes with arbitrary charges, determining the values of $h$ at which wall crossing occurs is difficult. 
However, following \cite{Manschot:2010qz}, one can focus on BPS states with a choice of charges belonging to sublattice $\tilde{\Gamma} =\{ M \xi_1 + N \xi_2 \,|\, M \geq 0, N \geq 0 \}$, where $(\xi_1 , \xi_2)$ are two non-parallel primitive vectors.\footnote{By primitive vectors, we mean vectors $\xi_i$ in the charge lattice such that $\xi/n$ does not belong to the charge lattice for any integer $n>1$. The case where the charges $\tilde{\Gamma}$ arrange themselves in the halo configurations described in section \ref{sec:halo_saddles} is referred to as a "semi-primitive" configuration; the more general configuration, which we review here, is called "non-primitive".} Assuming $\langle \xi_1 , \xi_2 \rangle<0$, one can define chambers $c^+$ and $c^-$ as
\be 
c^-: \qquad \langle \xi_1 , \xi_2 \rangle \Im(Z_{\xi_1} \bar{Z}_{\xi_2} ) >0 , 
\qquad 
c^+: \qquad \langle \xi_1 , \xi_2 \rangle \Im(Z_{\xi_1} \bar{Z}_{\xi_2} ) <0 
.
\label{eq:chamber-defintion}
\ee
We again start by reviewing the properties of Lorentzian extremal black hole solutions with such charges. 
In \cite{Manschot:2010qz}, it was shown that the Lorentzian multicentered bound states coming from sublattice $\tilde{\Gamma}$ only exist in the chamber $c^-$. 
In particular, the wall of marginal stability is located at points in moduli space where $\Im(Z_{\xi_1} \bar{Z}_{\xi_2} )=0$. 
This has the following implication.
For any black hole with monopole charge $\Gamma_i = n_i^1 \xi_1 + n_i^2 \xi_2 \in \tilde{\Gamma}$ belonging to the bound state, we have from explicit computation that
\be
\langle \Gamma_i , h \rangle \= \frac{2}{|Z_{\Gamma_{\text{total}}}|} (N_1 n_i^2 - N_2 n_i^1) 
\Im(Z_{\xi_1} \bar{Z}_{\xi_2} ) \,,
\ee
where we assumed the total charge $\Gamma_{\text{total}} = N_1 \xi_1 + N_2 \xi_2$ and we denoted $Z_{\Gamma_{\text{total}}} = \langle \Gamma_{\text{total}} , \Omega_\infty \rangle$.
Therefore, on the wall of marginal stability, all intersection products $\langle \Gamma_i , h \rangle = 0$ must vanish.

We can now proceed to analyze the condition for wall-crossing for the Euclidean finite temperature index saddles.  Let us start our analysis on the wall of marginal stability where $\langle \Gamma_i , h \rangle =0$. For this value of the scalars, there is an obvious finite temperature saddle: 
we can separate $|\xvec_i- \xvec_j|\to \infty$ $\forall \, i \neq j$, while keeping $|\xvec_i- \tilde \xvec_i|$ finite 
at the value 
\be 
\label{eq:distance-between-BHs-infty}
|\xvec_i- \tilde \xvec_i| \= \frac{\langle \Gamma_i, \delta_i \rangle}{\frac{\beta}{4\pi}+ \langle  \delta_i, h \rangle}\,,
\ee
which solves \eqref{eq:regularity_condition_as_real_equations1}--\eqref{eq:regularity_condition_as_real_equations4}. 
We can now slightly perturb this value of $h$ such that $\langle \Gamma_i , h \rangle = \epsilon_i$  in which case at least some of the $|\xvec_i- \xvec_j|$ 
values will need to be finite but large (of~$O(\epsilon^{-1})$) in order to solve the regularity constraints. 
We thus seek to check whether the regularity constraint can be solved at $O(\epsilon)$ in the chambers $c^\pm$. 
At $O(\epsilon)$, in  the regularity conditions   \eqref{eq:regularity_condition_as_real_equations1}  the terms containing $|\xvec_i- \xvec_j|$ cancel with those containing  $|\xvec_i- \tilde \xvec_j|$, $\forall\, i \neq j$, leaving only the term with $i=j$ in the second sum; 
similarly, in \eqref{eq:regularity_condition_as_real_equations2} the terms containing $|\tilde \xvec_i- \xvec_j|$ cancel with those containing  
$|\tilde \xvec_i- \tilde \xvec_j|$, $\forall\, i \neq j$, once again leaving only the term with $i=j$ in the first sum. 
Consequently, the first two regularity conditions  \eqref{eq:regularity_condition_as_real_equations1} and  \eqref{eq:regularity_condition_as_real_equations2} only imply that, at $O(\epsilon)$, the distance between each north-south pole pair is still given by~\eqref{eq:distance-between-BHs-infty}. Thus, we are solely left with solving \eqref{eq:regularity_condition_as_real_equations3} and  \eqref{eq:regularity_condition_as_real_equations4}. At $O(\epsilon)$, the two equations become equivalent and lead to
\begin{align}
\sum_{j=1 \atop j \neq i}^N \frac{\langle \Gamma_i, \Gamma_j \rangle}
{|\xvec_i - \xvec_j|}&\= 
- \epsilon_i
\,, \qquad 
i=1,\dots,N \,,
\label{eq:regularity_condition_close_to_wall}
\end{align}
where we have used that $\forall \, i \neq j$, $|\xvec_i- \xvec_j|= |\xvec_i-\tilde \xvec_j|= |\tilde \xvec_i- \xvec_j| =  |\tilde \xvec_i-\tilde \xvec_j|$ up to $O(\epsilon^2)$. 

Thus, even though we are working at an arbitrary finite temperature, \eqref{eq:regularity_condition_close_to_wall} precisely matches the zero temperature regularity conditions that  \cite{Manschot:2010qz} had to impose when going slightly away from the wall separating chambers $c^+$ and $c^-$. Since \eqref{eq:regularity_condition_close_to_wall} could not be solved in chamber $c^+$,\footnote{For example, the proof presented in \cite{Manschot:2010qz} that the integrability condition cannot be solved in the chamber $c^+$ relies on arranging the charge vectors $\Gamma_i$ in a clockwise order, $\langle \Gamma_i, \Gamma_j\rangle \geq 0,\, \forall\, i< j$. This consequently implies that in \eqref{eq:regularity_condition_close_to_wall}, $\epsilon_1 < 0$ while $\epsilon_{N} >0$. This constraints the imaginary part of $e^{-i \phi}Z_{\Gamma_i}$ ($e^{-i \phi}Z_{\Gamma_i}$) to be positive (negative) while the definition of the charge sublattice $\tilde{\Gamma}$ constrains the real part to be positive (positive). However, these conditions are incompatible with the definition \eqref{eq:chamber-defintion} of the chamber $c^+$. Following  \cite{Manschot:2010qz}, a slightly more complicated proof shows that solutions to \eqref{eq:regularity_condition_close_to_wall} indeed exist in the chamber $c^-$. } and solutions only exist in the $c^-$ chamber, finite temperature saddles (with $\xvec_i, \, \tilde \xvec_i \in \mathbb R^3$) also solely exist in the  $c^-$ chamber. Thus, even in this more generic case with $N>2$ with charges restricted to the sublattice $\tilde{\Gamma}$, we find that the walls of marginal stability occur at a value of $h$ that is (i) $\beta$-independent and (ii) agrees with the previously known results \cite{Manschot:2010qz}. It would be interesting to understand whether the mechanism for wall crossing, in which some of the black holes get separated by an infinite distance at the wall of marginal stability, is the same even for more generic choices of charges.

\section{Discussion}
\label{sec:discussion}

In this paper, we have constructed a family of multicentered supersymmetric black hole saddles that contribute to the gravitational index of 
$\mathcal{N}=2$ supergravity coupled to an arbitrary number of vector multiplets in four-dimensional asymptotically flat space. 
The new multicentered saddles satisfy the new form of attractor mechanism \cite{Boruch:2023gfn} -- while they depend sensitively on asymptotic boundary conditions for scalar fields and asymptotic temperature, 
their on-shell action gives only a trivial dependence on those parameters together with a temperature independent piece given by a sum over extremal entropies of the corresponding zero temperature black holes. 
The space of constructed solutions is continuous and parametrized by a set of Cartesian points subject to ``regularity conditions", a generalization of the zero temperature integrability conditions \cite{Bates:2003vx} which ensure smoothness of the underlying geometry. For the bound state of two black holes, we analyzed the regularity conditions in detail through a mixture of analytic and numerical methods and shown that resulting moduli space is given by $S^3 \times S^1$, in contrast to $S^2$ moduli space appearing in the extremal limit. We have further discussed some natural choices for symplectic form on the new moduli space and used it to analyze the vanishing of the saddles across walls of marginal stability. Below, we discuss some outstanding questions that deserve future investigation.

\paragraph{Further analysis of the space of solutions:}
In our detailed analysis of the finite temperature moduli space, we focused only on the two black hole bound state for a generic choice of charges. This led to the most constrained version of moduli space, due to all intersection products \eqref{eq:charge_intersections_assumptions} being nonvanishing. However, one can consider more specific choices of the charges for which some of these intersection products vanish, leading to a less constrained moduli space. It would be interesting to understand what are the possible bound state configurations contained in those moduli spaces.
Similarly, it would be interesting to understand different configurations appearing for more black holes. In particular, it seems that black hole halos of section \ref{sec:halo_saddles} could be a simple case to start with due to the explicit form of solution for distances given in \eqref{eq:halos_reg_sol1}--\eqref{eq:halos_reg_sol5}.

\paragraph{Quiver quantum mechanics:} An important role in understanding the zero temperature moduli space of multicentered BPS solutions is played by quiver quantum mechanics \cite{Denef:2002ru,deBoer:2008zn,Manschot:2011xc,Pioline:2013wta} which describes the system at small string coupling. The BPS moduli space of this theory is precisely described by integrability conditions \eqref{eq:extremal_Bates_Denef_integrability}, which therefore allows one to study the space of multicenter solutions directly from the quiver description. In the case of new finite temperature saddles of section \ref{sec:multicentered_Bates_Denef_at_finite_temperature}, the solution space is now instead described by regularity conditions \eqref{eq:multicentered_regularity_condition}. It would be interesting to understand if there exists a modified effective description for the new saddles at small string coupling. 

\paragraph{Positivity of the entropy function:} In Lorentzian signature, depending on the split of the charge total $\Gamma$ among the black hole configuration, the resulting solution could have an emblackening factor $\Sigma(\mathbf x)$ that becomes negative; 
in such cases, there are portions of the spacetime where the signature is changed and there exist closed timelike curves in the spacetime. 
For such a charge, splitting the solution is unphysical and is therefore not expected to contribute to the overall index. 
While discarding such solutions is sensible in Lorentzian signature, the Euclidean saddles studied in this paper already have a complexified metric for any generic charge splitting, and there is no widely accepted criterion for which saddles contribute to the GPI.\footnote{The Kontsevich-Siegel-Witten criterion \cite{Kontsevich:2021dmb,Witten:2021nzp,BenettiGenolini:2025jwe} is one such criterion. 
Yet, its regime of applicability is currently unknown as there are counterexamples where the criterion seemingly does not apply (see \cite{Bah:2022uyz, Chen:2023hra} for examples). }  
By analyzing the finite temperature saddles associated with the discarded Lorentzian solutions, one can hope to clarify the correct rules for which saddles contribute to the GPI. 

 \paragraph{Black rings, black lenses and multicenter $5d$ black holes.} The saddles for the $4d$ gravitational index can be uplifted to recover analogous contributions to the gravitational index of $5d$ theories. In \cite{Boruch:2025qdq}, such an uplift was performed to construct the single-center black hole and black string saddles for the $5d$ index. We plan to use the $4d$ multi-center saddles found in this paper to find a much richer set of $5d$ saddles. Depending on the choice of $4d$ charges, we expect to be able to recover the contributions of black rings \cite{Bandyopadhyay:2025jbc}, black lenses, and multicenter black holes to the $5d$ index \cite{wip}.

 \paragraph{One-loop determinants and the temperature independence of the index.} In sections~\ref{sec:Einstein_Maxwell} and~\ref{sec:sugra_with_vectors}, we have shown that the on-shell action of the multi-center saddles only has a trivial temperature dependence coming from the Boltzmann weight that BPS states have in the canonical ensemble. If the gravitational index indeed behaves like an ordinary Witten index in a supersymmetric quantum system, not only should the on-shell action have this trivial temperature dependence, but also all quantum corrections around such saddles should be completely temperature independent. In appendix \ref{sec:volume_of_moduli_space}, we have seen that in the simplest case of two black holes, the symplectic volume of the moduli space has a complicated temperature dependence. At first sight, this might seem in tension with the fact that all quantum corrections should be temperature independent. However, the index itself is not proportional to this symplectic volume. Rather, at each point in the moduli space, one has to compute the one- and higher-loop determinants to verify that the quantum corrections to the index are indeed temperature independent. We hope to study how this temperature independence arises in the near future.

\section*{Acknowledgments}
We want to especially thank Roberto Emparan for our collaboration on related directions. We also want to thank Ahmed Abdalla, Pawel Caputa, Matthew Heydeman, David Katona, Guanda Lin, Sebastian Mizera, Steve Shenker, Ashoke Sen, Douglas Stanford and Spencer Tamagni for helpful discussions. We are also grateful to Davide Cassani and Boris Pioline for comments about the draft.  LVI was supported by the DOE Early Career Award DE-SC0025522. GJT was supported by the University of Washington and the DOE award DE-SC0011637.
SM~acknowledges the support of the STFC grants ST/T000759/1,  ST/X000753/1.

\appendix

\section{Volume of the moduli space for two black holes}
\label{sec:volume_of_moduli_space}

In this appendix, we discuss volume of the finite temperature moduli space which we are able to compute explicitly in the case of two black holes for specific choices of the measure. Let us start with a brief review of the Lorentzian case. 
 
 In the extremal Lorentzian case, the moduli space of multicentered solutions carries a natural symplectic structure, given by 
\be 
\omega_{\text{extremal}} \= 
\frac{1}{4} 
\sum_{A<B}^{N} 
\langle
\Gamma_A, \Gamma_B
\rangle
\frac{\epsilon^{abc} \dd r_{AB}^a \wedge \dd r_{AB}^{b} 
r_{AB}^c}{|\vec{r}_{AB}|^3} \,,
\qquad
\Vec{r}_{AB} \equiv \xvec_A - \xvec_B \,,
\ee
which was determined in \cite{deBoer:2008zn} 
from the quiver quantum mechanics description at weak coupling. The above symplectic form is invariant under $SO(3)$ rotations, in the sense that the classical angular momentum \eqref{eq:extremal_angular_momentum}
of an extremal configuration is a moment map for the generator of rotations on the solution space 
\be 
\iota_X \omega_{\text{extremal}} \=
\dd J_z^{\text{extremal}} \,, 
\qquad 
X \= \sum_{A<B}^{N} \frac{\partial}{\partial \phi_{AB}} 
\,.
\ee
With this, it is then possible to study the volume of the extremal moduli space using localization. This was done in \cite{Manschot:2010qz,Manschot:2011xc} and led to results consistent with microscopic expectations (see \cite{Pioline:2013wta} for a review). 

In the finite temperature case, as of now, it is not clear to us how to study the solution space of complex multicentered index saddles using quiver quantum mechanics.\footnote{In the presence of a multicenter black hole configuration, each black hole has its own quantum system (in the holographic sense) and at the same time the multiple quantum systems are coupled by virtue of sharing the same ambient asymptotically flat space geometry. At zero temperatures each center corresponds to a pole of the harmonic functions specifying a solution. At finite temperatures some pairs of poles correspond to special points within the same horizon while other pairs correspond to different black holes. Therefore all pairs of poles cannot be on equal footings in the description of the system in terms of quantum mechanics. We believe this explains why the measure derived from quiver quantum mechanics does not strictly apply at finite temperatures. This is clear from the fact that the measure leads to a temperature-dependent volume.} We can, however, write down a natural guess for the symplectic form based purely on the requirement of $SO(3)$ invariance. Requiring that the solution space of the attractor saddles satisfies 
\be 
\iota_X \omega_\beta \=
\dd J_z \,, 
\qquad 
X \= \sum_{A<B}^{2N} \frac{\partial}{\partial \phi_{AB}} 
\,,
\ee
with the finite temperature moment map given by 
\be 
\label{eq:angular-momentum-expression-2}
J_z \= 
\frac{1}{2} \sum_{A<B}^{2N} 
\langle
\gamma_A, \gamma_B
\rangle
\cos \theta_{AB}
- \frac{\beta}{8\pi \i} \sum_i^{N} (z_i -z_{\overline{i}})
\,,
\ee
implies the following symplectic form
\begin{align}
\label{eq:symplectic-form-NBH}
\omega_\beta \= 
\frac{1}{4} 
\sum_{A<B}^{2N} 
\langle
\gamma_A, \gamma_B
\rangle
\frac{\epsilon^{abc} \dd r_{AB}^a \wedge \dd r_{AB}^{b} 
r_{AB}^c}{|r_{AB}|^3}
\; + \; \frac{\beta}{8\pi \i}
\sum_i^{n_{\text{BH}}}
\dd (z_i - z_{\overline{i}}) \wedge \dd \phi_{i\overline{i}} 
\end{align}
where we introduced collective variables $\gamma_A \equiv (\gamma_i , \tgamma_j)_{i,j=1\dots N}$ and $\xvec_A = (\xvec_i , \txvec_j)_{i,j=1\dots N}$. Note that the constructed symplectic form is not real. This is related to the fact that saddles we work with are intrinsically complex. However, one could hope that it can still be used to extract a real volume integral due to the existence of complex conjugate saddles discussed in figure \ref{fig:complex_conjugate_saddles}. This indeed can happen, as we verify on the two black hole example below. 

Restricting ourselves now to finite temperature bound state configurations of section \ref{sec:2BH_moduli_space}, the solution space is four dimensional and correspondingly the moduli space volume is given by
\be 
\text{Vol}(\mathcal{M}_{2\text{BH}}) 
\= \int_{\mathcal{M}_{2\text{BH}}} \dd^4 x \, \text{Pf}(\omega_\beta)
\= \int_{\mathcal{M}_{2\text{BH}}} \frac{\omega_\beta^2}{2!} \, .
\ee
The solution space $\mathcal{M}_{2\text{BH}}$ can be parametrized by three Euler angles $(\theta,\phi,\sigma)$ and a distance $x_{12}=|\xvec_1 - \xvec_2|$ as in figure \ref{fig:general_S03_config}. Since we know the edges of moduli space \eqref{eq:edges_analytic_perturbative_LEFT},\eqref{eq:edges_analytic_perturbative_RIGHT}, we can simply perform the integral directly. 
The explicit expression for the Pfaffian in the new parametrization is quite complicated. However, to prove that the final volume is real, it is enough to notice that the imaginary part of the Pfaffian has a simple periodic dependence on $\sigma$, and therefore after integrating over $\sigma$ we cancel out the imaginary parts 
\be 
\int_0^{2\pi} \dd \sigma \, \text{Pf}(\omega_\beta) 
\in \mathbb{R} \,. 
\ee
The cancellation of the imaginary parts in $\sigma$ integral happens precisely because complex conjugate saddles are already contained in the moduli space and are related to each other in pairs through transformation $\sigma \to \sigma + \pi$ (see figure \ref{fig:complex_conjugate_saddles}). The remaining three integrals can be performed analytically in perturbation theory for large $\beta$. The leading part of the answer is given by
\begin{align}
\text{Vol}(\mathcal{M}_{2\text{BH}}) 
\= \frac{64\pi^4}{\beta^2} 
\frac{\Gidi \Gzdz 
\left(
\hdi \hdz \GiGz + \hGi^2 \didz 
\right)
}{\GiGz} 
+ O (\beta^{-3})
\,.
\end{align}

\bibliographystyle{utphys2}
{\small \bibliography{Biblio}{}}

\end{document}